\documentclass[10pt,journal]{IEEEtran}

\usepackage{cite}

\ifCLASSINFOpdf
  \usepackage[pdftex]{graphicx}
  \DeclareGraphicsExtensions{.pdf,.jpeg,.png}
\else
\fi
\usepackage[cmex10]{amsmath}
\usepackage{amsfonts,amssymb,amsthm,mdwmath,bbm}
\usepackage{array}

\ifCLASSOPTIONcompsoc
  \usepackage[caption=false,font=normalsize,labelfont=sf,textfont=sf]{subfig}
\else
  \usepackage[caption=false,font=footnotesize]{subfig}
\fi

\usepackage{stfloats}
\usepackage{url}
\usepackage{color}
\usepackage{acronym}
\usepackage{enumitem}

\begin{document}
\acrodef{PPP}[PPP]{Poisson point process}
\acrodef{PGFL}[PGFL]{probability generating functional}
\acrodef{CDF}[CDF]{cumulative distribution function}
\acrodef{PDF}[PDF]{probability distribution function}
\acrodef{PCF}[PCF]{pair correlation function}
\acrodef{MoM}[MoM]{method of moments}
\acrodef{MLE}[MLE]{maximum likelihood estimate}
\acrodef{RV}[RV]{random variable}
\acrodef{V2V}[V2V]{vehicle-to-vehicle}
\acrodef{SIR}[SIR]{signal-to-interference ratio}
\acrodef{1D}[1D]{one-dimensional}
\acrodef{VANET}[VANET]{vehicular ad hoc network}
\acrodef{LT}[LT]{Laplace Transform}

\title{Outage in Motorway Multi-Lane VANETs with Hardcore Headway Distance Using Synthetic Traces}

\author{Konstantinos Koufos and Carl P. Dettmann 
\thanks{K.~Koufos and C.P.~Dettmann are with the School of Mathematics, University of Bristol, BS8 1TW, Bristol, UK. \{K.Koufos, Carl.Dettmann\}@bristol.ac.uk} \protect \\ 
}

\maketitle

\begin{abstract}
In this paper we analyze synthetic mobility traces generated for three-lane unidirectional motorway traffic to find that the locations of vehicles along a lane are better modeled by a hardcore point process instead of the widely-accepted Poisson point process (PPP). In order to capture the repulsion between successive vehicles while maintaining a level of analytical tractability, we make a simple extension to PPP: We model the inter-vehicle distance along a lane equal to the sum of a constant hardcore distance and an exponentially distributed random variable. We calculate the J-function and the Ripley's K-function for this hardcore point process. We fit its parameters to the available traces, and we illustrate that the higher the average speed along a lane, the more prominent the hardcore component becomes. In addition, we consider a transmitter-receiver link on the same lane, and we generate simple formulae for the moments of interference under reduced Palm measure for that lane, and without conditioning for other lanes. We illustrate that under Rayleigh fading a shifted-gamma approximation for the distribution of interference per lane provides a very good fit to the simulated outage probability using the synthetic traces, while the fit using the PPP is poor. 
\end{abstract}


\section{Introduction}
The world-wide PHY/MAC layer standard forming the basis of \ac{V2V} communication is the IEEE 802.11p in the 5.9 GHz band~\cite{FCC2003}. With the ongoing roll-out of 5G network deployments, other forms of vehicular communication, e.g., vehicle-to-infrastructure, will also be available in order to enhance traffic efficiency, safety and meet the increasing demands for high throughput and low latency at the vehicles. The \ac{V2V} communication will be secured in the cellular band too, complementing IEEE 802.11p, because it is of paramount importance for certain applications, e.g., autonomous driving, platooning, velocity and brake control.  

Applications and protocols for \ac{V2V} communication have been extensively investigated during the past two decades~\cite{Kihl2008,Willke2009}. The cost of deploying large scale testbeds is high, and the proposed solutions had been mostly assessed using computer simulations~\cite{Harri2009}. Modern simulators can include street maps, and realistic micro-mobility behavior, e.g., lane changing, acceleration/deceleration and car-following patterns~\cite{SUMO}. In addition, they can be calibrated with measurements for macroscopic features like intensity and average speed of vehicles, giving rise to synthetic mobility traces. The traces available in~\cite{Fiore2014, Fiore2015} are valuable, because they can be used to validate the performance obtained with simplified deployment models, as we will do in this paper. 

With the recent advent of wireless networks with irregular structure, e.g., small cells, sensor and wireless ad hoc networks, point processes have been employed to model the locations of network elements and investigate their performance~\cite{Haenggi2009a}. In vehicular networks, the spatial model can be divided into two components: the road infrastructure and the deployment of vehicles along a road. Instead of running time-consuming simulations, the analysis with point processes can give a quick insight into the impact of various parameters on the properties of the network. The analytical results should be trusted only if the adopted processes are realistic. 

Modeling the road infrastructure and the distribution of headways has long been a subject studied in transportation research. The adopted models are often complicated: Random tessellations have been fitted to real data of inter-city main roads and side streets, minimizing a distance metric~\cite{Gloaguen2006}. Empirical studies revealed that the distribution of time headway follows closely the log-normal distribution under free flow~\cite{Daou1966} and the log-logistic distribution under congestion~\cite{Yin2009}. The distribution of vehicles may also be modeled with a two-dimensional point process. The Thomas, Mat{\`e}rn cluster and log-Gaussian Cox processes fit well real snapshots of taxis, independently of the regularity of urban street layouts~\cite{Haenggi2019}. Despite their impressive accuracy, these models seem quite complex to incorporate into the performance evaluation of \acp{VANET}. In order to balance between accuracy and tractability, the Poisson line process can be used to model roads with random orientation, coupled with \ac{1D} \ac{PPP} for the locations of vehicles along each road~\cite{Dhillon2018,Baccelli2018}. Under these assumptions, the distribution of vehicles becomes a Cox process, and the coverage probability of a typical vehicle can be derived in semi-closed form~\cite[Theorem 1]{Dhillon2018}. Therein, the conflicting effect of road intensity (higher intensity increases the interference level) and vehicle intensity (higher intensity increases the average link gain) is also demonstrated. Furthermore, the Manhattan Poisson line process can model a regular layout of streets, with the empty space filled-in with objects to resemble urban districts~\cite{Baccelli2015}. Near intersections, the packet reception probability decreases because there is dominant interference both from horizontal and vertical streets~\cite{Steinmetz2015}. 

Inter-city motorway traffic does not require a complex model for the road network. A superposition of \ac{1D} point processes should suffice to model the locations of vehicles along multi-lane motorways. Not surprisingly, the \ac{PPP} has been widely adopted in these scenarios. Due to its simplicity, it has been used in the performance evaluation of complex communication protocols with multi-hop interference~\cite{Hu2016} and channel access schemes with collision-avoidance~\cite{Alouini2016}. Unfortunately, the \ac{PPP} will be accurate only under certain conditions. For instance, its independence assumption may not hold near traffic lights due to clustering~\cite{Polak2011}. Also, in high-speed motorways, the drivers maintain a safety distance from the vehicle ahead, depending on their speed and reaction time~\cite{Cowan1975}. The study in~\cite{Guo2018} shows that for a Poisson flow of vehicles entering a road, the headway distance follows the exponential distribution in the steady state, under the assumption that the vehicles select in the entrance of the road their speed from a common \ac{PDF}. The study in~\cite{Hourani2018} establishes the suitability of \ac{1D} \ac{PPP} for low transmission probability per vehicle. This assumption might be true with the underlying automotive radar application, where each vehicle sends a short pulse and waits for the response during the duty cycle. Intuitively, under strong thinning, the interference field due to a lattice converges to that due to a \ac{PPP} of equal intensity, and the \ac{PPP} becomes a valid model for target detection. Diverting from the \ac{PPP}'s independence assumption adds very high complexity in the performance evaluation. The lifetime for a link with log-normally distributed headway distances is studied in~\cite{Yan2011}, but the impact of interference is neglected. The bit error probability and channel capacity are studied in~\cite{Win2013} with a non-uniform intensity measure modeling clusters of vehicles, but the interference is neglected there too. 

In~\cite{Koufos2018,Koufos2019}, we have taken a step away from the \ac{PPP}, modeling the headway distance equal to the sum of a constant hardcore distance and an exponentially distributed \ac{RV}. The hardcore distance models the minimum spacing between successive vehicles along a lane. This model is known in transportation research as the Cowan M2~\cite{Cowan1975}. Its correlation properties have been studied since the early 1950's in statistical mechanics, under the name of radial distribution function for hard spheres, where the spheres are the particles of \ac{1D} fluids~\cite{Salsburg1953}. In~\cite{Koufos2018,Koufos2019} we simplified its \ac{PCF} keeping only short-range correlations, and we generated simple expressions for the variance of interference, the skewness and the probability of outage at the origin. 

In the current paper, we extend the interference and outage models of~\cite{Koufos2018,Koufos2019} to multi-lane \acp{VANET}, super-imposing independent Cowan M2 models for each lane. Most importantly, we analyze synthetic traces of motorway traffic~\cite{Fiore2014,Fiore2015} and validate our models against spatial statistics. Furthermore, we devise simple models for the probability of outage under multi-lane interference, and we illustrate that the model predictions using Cowan M2 agrees well with the empirical probability of outage simulated using the traces. On the other hand, the performance predictions using the PPP fail. To the best of our knowledge, the interference performance of multi-lane \acp{VANET} has not been assessed before with a repulsive point process for the vehicles. In addition, the existing interference models, based on the \ac{PPP}, have not been validated against real vehicular traces.

Empirical data has already been used in the performance evaluation of cellular wireless networks~\cite{Guo2013}. For vehicular networks, the study in~\cite{Haenggi2019} uses real locations of taxis and points out the dependency in their locations through sampling. It shows that the Log Gaussian Cox process minimizes the contrast to the Ripley's K-function and to the connection probability. In~\cite{Fiore2014, Fiore2015}, the authors processed measurement data about the intensity and speed of vehicles per lane as they pass a point of a motorway. They generated synthetic traces to investigate the topology of traffic, and highlight the impact of communication range on full connectivity, however, without considering interference. We would like to see whether a simple enhancement to \ac{PPP}, i.e., Cowan M2, can predict well the empirical (using the traces of~\cite{Fiore2014, Fiore2015}) distribution of \ac{SIR} in motorway \acp{VANET}. The contributions of our work are: 
\begin{itemize}[leftmargin=*]
\item We analyze the synthetic traces~\cite{Fiore2014, Fiore2015}, and we illustrate that the envelopes of the J- and the L- functions for small distances indicate repulsion. The leftmost lane which is characterized by the highest average speed of vehicles experiences the highest repulsion. This suggests that Cowan M2 might be a better model than \ac{PPP}, because the repulsion can be captured through the hardcore distance. 
\item We calculate the J-function of the hardcore process in closed-form and the Ripleys's K-function as a finite sum. These functions can be used for fitting the parameters of the hardcore process to the snapshot. We illustrate that the J-function of the hardcore process fits well the empirical J-function for small distances. 
\item We generate simple but accurate approximations for the mean, the variance and the skewness of interference under reduced Palm for the hardcore point process. Then, we approximate the \ac{LT} of interference from the lane containing the transmitter-receiver link. Note that the moments of interference derived in~\cite{Koufos2018,Koufos2019} are without conditioning and thus, they will be used to approximate the \ac{LT} of interference from other lanes.
\item  We illustrate that the hardcore process coupled with a shifted-gamma approximation for the distribution of interference per lane fits well the empirical outage probability generated, while the fit using \ac{PPP} is poor. We assess the approximation accuracy by goodness-of-fit metrics. 
\end{itemize}

The remainder of this paper is organized as follows: In Section~\ref{sec:Model}, we introduce the system model for single lane. In Section~\ref{sec:Summary}, we calculate the summary statistics for the hardcore point process. In Section~\ref{sec:Validation} we show how to fit the parameters of the hardcore point process to the synthetic traces. In Section~\ref{sec:MeanVariance}, we approximate the first three moments of interference under reduced Palm, and we fit a shifted-gamma approximation for the distribution of interference. In Section~\ref{sec:MultiLane}, we extend the model to multiple lanes, and in Section~\ref{sec:MultiLaneSynthetic} we validate it against synthetic traces. In Section~\ref{sec:Conclusions}, we summarize the main results of this study. 

\section{System model}
\label{sec:Model}
We consider a \ac{1D} point process of vehicles $\Phi$, where the inter-vehicle distance follows the shifted-exponential \ac{PDF}. The shift is denoted by $c\!>\!0$ and the rate by $\mu\!>\! 0$. The intensity $\lambda$ of vehicles is calculated from $\lambda^{-1}\!=\!c\!+\!\mu^{-1}$, or equivalently  $\lambda\!=\!\frac{\mu}{1+\mu c}$. The joint probability that there are two vehicles at $x$ and $y\!>\!x$, is $\rho^{\left(2\right)}\!\left(y,x\right)\!=\!\sum\nolimits_{k=1}^\infty \rho_k^{\left(2\right)}\!\left(y,x\right)$, where 
\begin{equation}
\label{eq:rho2}
\rho_k^{\left(2\right)}\!\!\left(y,x\right) \!\!=\!\! \Bigg\{ \!\!\!\! \begin{array}{ccl}\lambda \! \sum\limits_{j=1}^k \!\! \frac{\mu^j \left(y-x-jc\right)^{j-1}}{\Gamma\left(j\right) e^{\mu\left(y-x-jc\right)}}, \!\!\!\!\!\!& &\!\!\!\!\!\! y\!\in\!\left(x\!\!+\!kc, x\!\!+\!\left(k\!+\!1\right)\!c\right) \\ 0, \!\!\!\!\!\!& &\!\!\!\!\!\! \text{otherwise}, \end{array} 
\end{equation}
$k\!\geq\! 1$ and $\Gamma\!\left(j\right)\!=\!\left(j\!-\!1\right)!$~\cite[equation~(32)]{Salsburg1953}.
\begin{figure}[!t]
 \centering
  \includegraphics[width=3in]{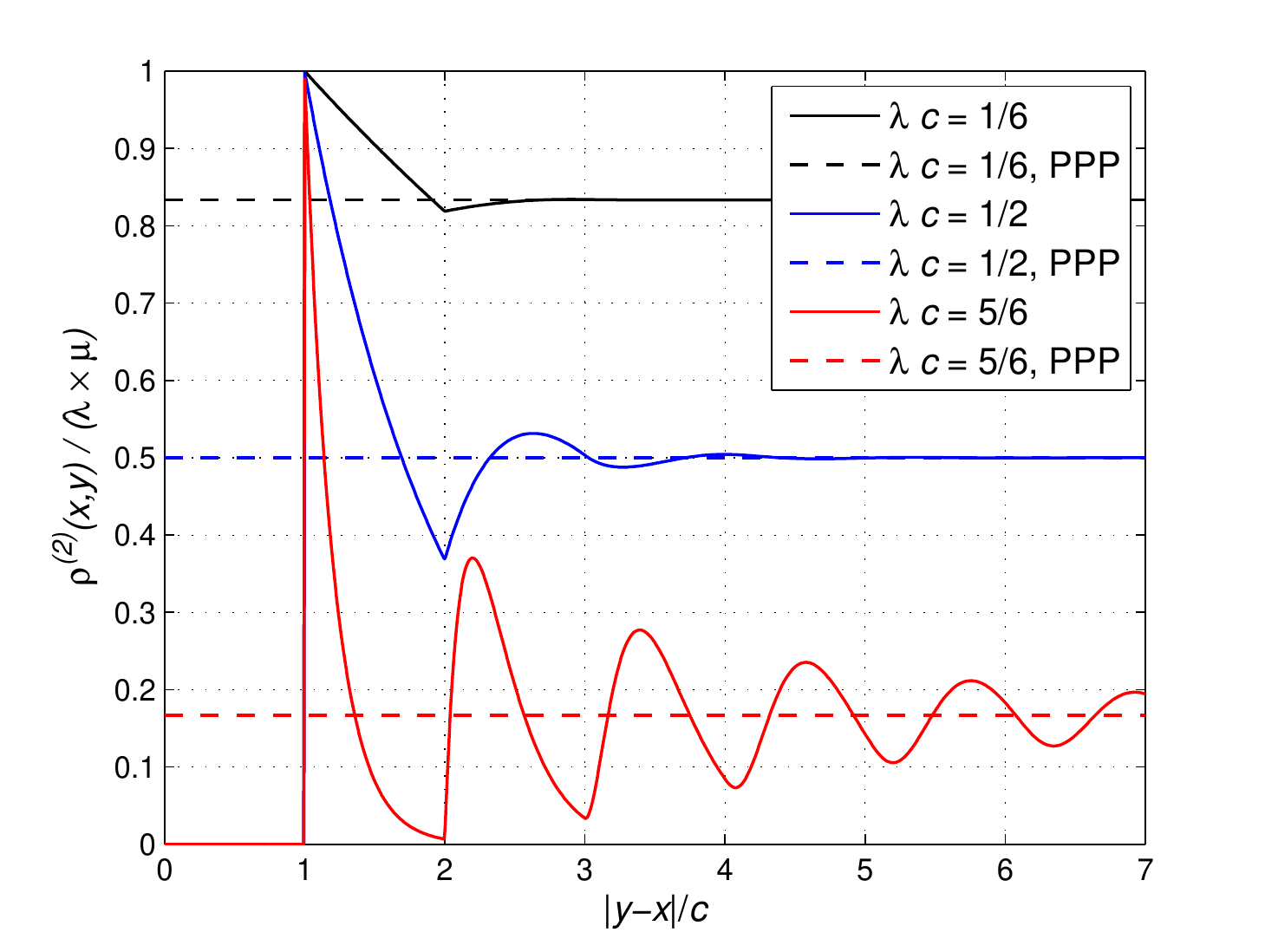}
 \caption{Normalized \ac{PCF}  $\rho^{\left(2\right)}\!\left(x,y\right)\left(\lambda \mu\right)^{-1}$ with respect to the normalized distance $|y-x|c^{-1}$.  The dashed horizontal lines correspond to $\rho^{\left(2\right)}\!\left(x,y\right)\!=\!\lambda^2$, or, $\rho^{\left(2\right)}\!\left(x,y\right)\left(\lambda \mu\right)^{-1}\!=\!1\!-\!\lambda c$~\cite[Fig.2]{Koufos2018}.}
 \label{fig:CorrFunc}
\end{figure}

The \ac{PCF}, $\rho^{\left(2\right)}\!\left(y,x\right)$, is depicted in Fig.~\ref{fig:CorrFunc}. For small hardcore distance $c$ as compared to the mean inter-vehicle distance $\lambda^{-1}$, the \ac{PCF} converges quickly to $\lambda^2$, which is the \ac{PCF} of a \ac{PPP} of intensity $\lambda$. Indeed, the locations of vehicles become  uncorrelated at few multiples of $c$ for $\lambda c \!\ll\! 1$. 

The higher-order correlations are naturally more complicated than the \ac{PCF}. For $n$ ordered points, $x_1,x_2,\ldots x_n$, the $n-$th order correlation is  $\rho^{\left(n\right)}\!\left(x_1,x_2,\ldots x_n \right)\!=\! \frac{1}{\lambda^{n-2}}\prod_{i=1}^{n-1}\rho^{\left(2\right)}\!\left(x_{i+1}\!-\!x_i \right)$~\cite[equation~(27)]{Salsburg1953}. For instance, the third-order correlation describing the probability to find a triple of distinct vehicles at $x,y$ and $z$, is
\begin{equation}
\label{eq:rho3}
\rho^{\left(3\right)}\!\left(x,y,z\right) = \frac{1}{\lambda} \,  \rho^{\left(2\right)}\!\left(x,y\right) \rho^{\left(2\right)}\!\left(y,z\right), \, x\!>\!y\!>\!z.
\end{equation}

\begin{figure}[!t]
 \centering
  \includegraphics[width=\columnwidth]{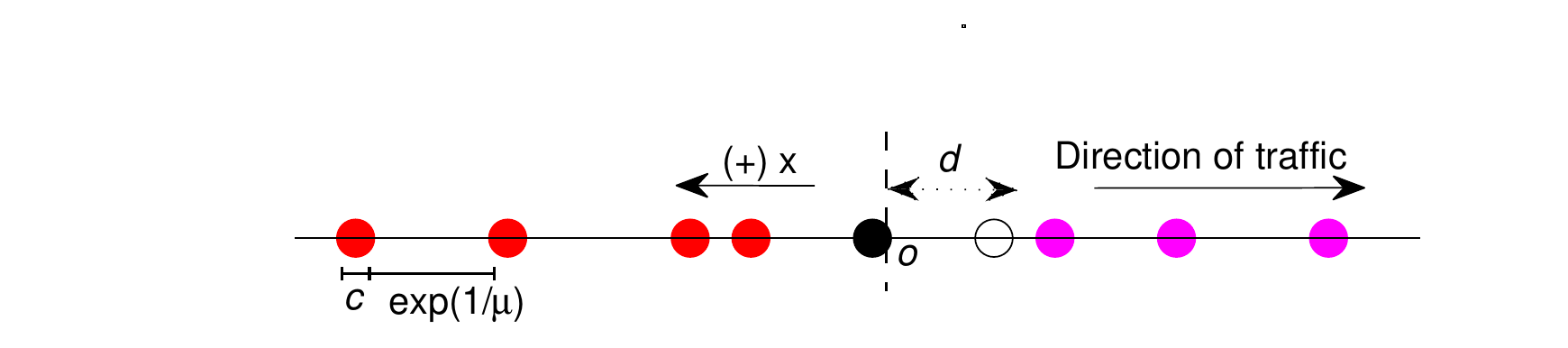}
 \caption{The average length of a vehicle and the safety distance a driver maintains from the vehicle ahead is modeled as an impenetrable disk of diameter $c$. The antenna of a vehicle is placed at the right side of the disk. A transmitter (black disk) is conditioned at the origin and paired with the receiver (hollow  disk) at $x\!=\!-d$. All the vehicles move rightwards. The vehicles behind the transmitter (red disks) are potential sources of interference. The vehicles in front of the receiver (magenta disks) are potential sources of interference too, but their transmission is attenuated due to the antenna backlobe radiation.}
 \label{fig:SystModelBeam}
\end{figure}

We condition the location of a transmitter at the origin, without loss of generality. The associated receiver is at distance $d$ away, see  Fig.~\ref{fig:SystModelBeam}. Inter-vehicle communication between successive vehicles at the same lane of a motorway would find possible applications in velocity, brake and adaptive cruise control. The locations of vehicles behind the transmitter and in front of the receiver follow the point process $\Phi$. The distance $d$ follows the shifted-exponential distribution too, with shift $c$ and rate $\mu$. Given $d$, the distance-based useful signal level at the receiver is denoted by $P_r\!\left(d\right)$. We assume that the transmissions of the vehicles in front of the receiver are attenuated by a positive factor $g\!<\!1$. 

The transmit power level is normalized to unity for all the vehicles. The propagation pathloss exponent is denoted by $\eta\!>\! 1$. The distance-based pathloss is $g\!\left(r\right)\!=\!r^{-\eta}$. The fading power level over the interfering links, $h$, and over the transmitter-receiver link, $h_t$, is exponential (Rayleigh distribution for the fading amplitudes) with mean unity. The fading is independent and identically distributed over different links. Each interferer is active with probability $\xi$. 

\section{Spatial statistics}
\label{sec:Summary}
A classic measure characterizing the behavior for a set of points (repulsion, clustering or complete randomness) is the J-function at distance $r$, ${\text{J}}\!\left(r\right)\!\triangleq\! \frac{1-G\left(r\right)}{1-F\left(r\right)}$~\cite[Chapter 2.8]{Haenggi2013}. It is the ratio of two complementary \acp{CDF}: the nearest-neighbor distance \ac{CDF} $G\!\left(r\right)$ divided by the contact \ac{CDF} $F\!\left(r\right)$. The nearest-neighbor distance is the distance between a point $x\!\in\!\Phi$ and its nearest neighbor $\min\nolimits_y \lVert x\!-\!y \rVert, y\!\in\!\Phi \backslash  \left\{x\right\}$. The contact distance is the distance between a reference location $u$ and the nearest point $x\!\in\!\Phi$, i.e., $\min\nolimits_{x\in\Phi}\lVert u\!-\!x \rVert$. For a \ac{PPP}, ${\text{J}}\!\left(r\right)\!=\!1$, e.g., in \ac{1D} both $G\!\left(r\right)$ and $F\!\left(r\right)$ are exponential with rate twice the intensity. For a repulsive point process with hardcore distance $c$, $G\!\left(r\right)\!=\!0, r\!\leq\!c$ and $F\!\left(r\right)\!\geq\!0, r\!\leq\!c$, resulting in ${\text{J}}\!\left(r\right)\!\geq\!1, r\!\leq\!c$. On the other hand, ${\text{J}}\!\left(r\right)\!<\!1$ is associated with clustering. 

The function $G\!\left(r\right)$ for the hardcore point process $\Phi$ introduced in Section~\ref{sec:Model} is shifted-exponential, $G\!\left(r\right)\!=\!1\!-\!e^{-2\mu\left(r-c\right)}, r\!\geq\! c$. Its contact \ac{CDF} $F\!\left(r\right)$ has been derived in~\cite[Section III]{Koufos2019}, and it has a piecewise form with breakpoint at $\frac{c}{2}$. After substituting $G$ and $F$, into ${\text{J}}$, we have 
\begin{equation}
\label{eq:Jfunction}
{\text{J}}\!\left(r\right)=\Bigg\{ \begin{array}{lcl} \left(1\!-\!2\lambda r\right)^{-1},& & r\!\leq\!\frac{c}{2} \\ \left(1-\lambda c\right)^{-1} e^{\mu\left(2r-c\right)}, & & \frac{c}{2}\!<\!r\!\leq\!c \\ 
\left(1-\lambda c\right)^{-1} e^{\mu c}, & & r\!>\! c. \end{array}
\end{equation}

It is straightforward to see that ${\text{J}}\!\left(r\right)\!>\!1 \, \forall r\!>\! 0$ since $\lambda c\!<\! 1$. The J-function increases up to $r\!=\!c$, and then remains constant, indicating the repulsion for distances $r\!\leq\!c$. The amplitude of the J-function is an indicator of the repulsiveness and can be used to compare two hardcore point processes at a given distance. The J-function uses empty-space distributions and thus, it is not very informative about the long-range behavior of the process. Another metric directly targeting the second-order properties is the Ripley's K-function~\cite[Chapter 2.5]{Baddeley2007}. The K-function counts the mean number of points within a distance $r$ from a point of the process (without counting this point), and normalizes the outcome with the intensity $\lambda$:  ${\text{K}}\!\left(r\right)\!\triangleq\! \frac{1}{\lambda}\mathbb{E}\left\{\Phi\left(B\!\left(x,r\right)\right)\!-\!1\right\}$, where $B\!\left(x,r\right)$ is a ball with radius $r$ centered at $x\!\in\!\Phi$. For \ac{1D} \ac{PPP} of intensity $\lambda$, the mean number of points within $B\!\left(x,r\right)$ and excluding $x$ is $2\lambda r$. Therefore normalizing the K-function by two can be used to distinguish between repulsion and clustering in \ac{1D}. For instance, ${\text{L}}\!\left(r\right)\!\triangleq\!\frac{{\text{K}}\!\left(r\right)}{2}\!<\!r$ means that the point process contains fewer points as compared to the \ac{PPP} of equal intensity within $B\!\left(x,r\right)$ indicating repulsion. The normalized Ripley's K-function is referred to as the Besag L-function. The K-function is also the integral of $\frac{1}{\lambda}\rho^{\left(2\right)}\!\left(y,x\right)$ within the area centered at $x\!\in\!\Phi$ and extending up to distance $r$ from $x$~\cite[Chapter 2.5]{Baddeley2007}. In our system and notational setup we get 
\[
\begin{array}{ccl}
{\text{K}}\!\left(r\right) \!\!\!\!\! &=& \!\!\!\!\! \displaystyle \frac{2}{\lambda^2} \int_0^r  \rho^{\left(2\right)}\!\left(r\right){\rm d}r = \frac{2}{\lambda^2} \int_0^r \sum\limits_{k=1}^{\left\lfloor{r/c}\right\rfloor} \rho_k^{\left(2\right)}\!\left(r\right) {\rm d}r \\ \!\!\!\!\! &=& \!\!\!\!\! \displaystyle \frac{2}{\lambda}\!\int_0^r \! \sum\limits_{k=1}^{\left\lfloor{r/c}\right\rfloor} \! \sum\limits_{j=1}^{k} \! \frac{\mu^j \left(r\!-\!jc\right)^{j-1}\mathbbm{1}\left(kc\!\leq\!r\!\leq \left(k\!+\!1\right)c\right)}{\Gamma\left(j\right) e^{\mu\left(r-jc\right)}} {\rm d}r, 
\end{array}
\]
where $x\!\equiv\!o$, $\mathbbm{1}$ is the indicator function, the factor two accounts for the negative half-axis, and also note that in~\eqref{eq:rho2}, the \ac{PCF} $\rho^{\left(2\right)}\!\left(y,x\right)$ is defined as the joint probability to find a pair of vehicles at $x,y$, thereby we have to divide by $\lambda$ since the K-function conditions on the location of a vehicle at $x$. 

The above expression can be simplified by exchanging the orders of summation (over $k$) and integration. 
\[
\begin{array}{ccl}
{\text{K}}\!\left(r\right) \!\!\!\!\! &=& \!\!\!\!\! \displaystyle \frac{2}{\lambda} \sum\limits_{k=1}^{\left\lfloor{r/c}\right\rfloor} \!\! \int_0^r \!\! \sum\limits_{j=1}^{k} \!\! \frac{\mu^j \! \left(r\!-\!jc\right)^{j-1}\mathbbm{1}\left(kc\!\leq\!r\!\leq \left(k\!+\!1\right)c\right)}{\Gamma\left(j\right) e^{\mu\left(r-jc\right)}}{\rm d}r \\ \!\!\!\!\! &=& \!\!\!\!\! \displaystyle \frac{2}{\lambda} \sum\limits_{k=1}^{\left\lfloor{r/c}\right\rfloor} \int_{kc}^r \frac{\mu^k \left(r-kc\right)^{k-1}}{\Gamma\left(k\right) e^{\mu\left(r-kc\right)}} {\rm d}r. 
\end{array}
\] 

\noindent 
After carrying out the integration in terms of $r$ we get 
\begin{equation}
\label{eq:Kfunction}
{\text{K}}\!\left(r\right) = \frac{2}{\lambda} \sum\nolimits_{k=1}^{\left\lfloor{r/c}\right\rfloor} \left(1\!-\!\frac{\Gamma\!\left(k,\mu\left( r\!-\!c k \right)\right)}{\Gamma\!\left(k\right)}\right),
\end{equation}
where $\Gamma\!\left(a,x\right)\!=\!\int_x^\infty \!\frac{t^{a-1}}{e^t}{\rm d}t$ is the incomplete Gamma function.
\begin{figure}[!t]
 \centering 
  \includegraphics[width=3in]{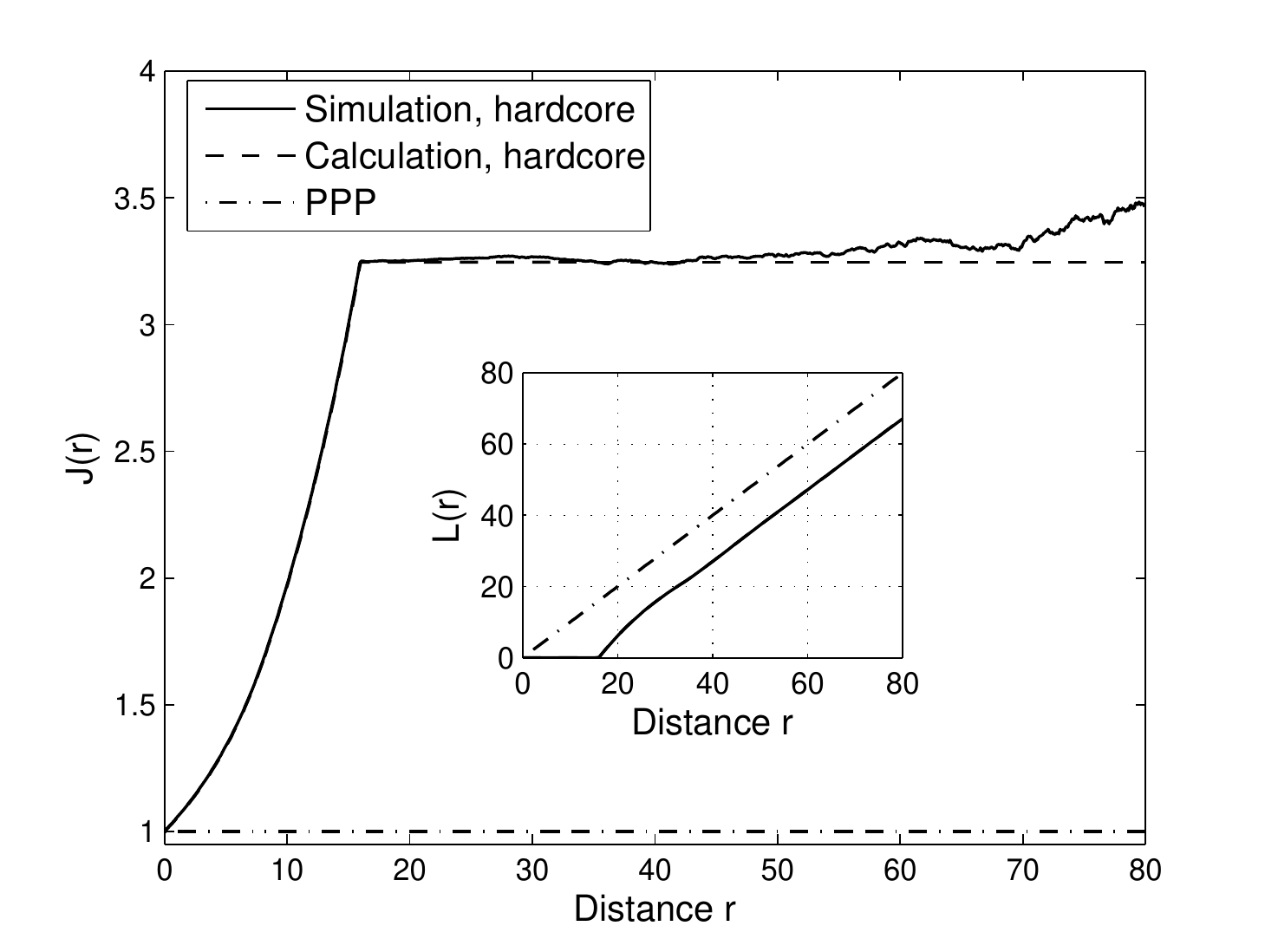}
 \caption{Spatial statistics for the hardcore process. Intensity $\lambda\!=\!0.025 {\text{m}}^{-1}$ and hardcore distance $c\!=\! 16$ m, yielding $\lambda c\!=\!0.4$. $10^6$ simulations within a line segment of $10$ km. In each simulation, we measure the nearest-neighbor distance for a point of the process, and the contact distance to the origin. The calculation is given in~\eqref{eq:Jfunction}. In order to generate the inset, in each simulation we select a point of the process close to the origin and we count the points of the process within distance $r$ from that point. Then we normalize by $2\lambda$. The calculation is given in~\eqref{eq:Kfunction} after scaling by half. The calculated curve for the L-function (not visible) overlaps with the simulations.}
\label{fig:JfunctionHC}
\end{figure} 

Example illustrations for the ${\text{J}}\!\left(r\right)$ and ${\text{L}}\!\left(r\right)$ are depicted in Fig.~\ref{fig:JfunctionHC}. Both functions capture the repulsion of $\Phi$ and quantify the hardcore distance $c$ by visual inspection. For a realization of $\Phi$ within a finite domain, the evaluation of ${\text{F}}\!\left(r\right)$ is constrained by the half of the maximum inter-point distance, denoted by $r_f$, while the evaluation of ${\text{G}}\!\left(r\right)$ is constrained by the maximum nearest-neighbor distance $r_g$. Both $r_f, r_g$ increase logarithmically with the system size. In most realizations of $\Phi$ we have $r_f\!<\!r_g$, which means that the function ${\text{J}}\!\left(r\right)$ will start to take very large values close to $r_f$. Because of that, in Fig.~\ref{fig:JfunctionHC}, the simulated average starts to rise for $r\!>\! 70$. In the inset of Fig.~\ref{fig:JfunctionHC}, the slope of the L-function for $r\!>\! 50$ m becomes practically unity, indicating that the correlation for distance separations larger than approximately $3c$ is not too strong for $\lambda c\!=\!0.4$. This remark is in accordance with the 'blue curve' in Fig.~\ref{fig:CorrFunc}, and it cannot be deduced from the graph of J-function. Next, we shall generate the summary statistics of synthetic traces, and study whether the hardcore process $\Phi$ can capture them better than the \ac{PPP}. 

\section{Validating the model with synthetic traces}
\label{sec:Validation}
The studies in~\cite{Fiore2014,Fiore2015} use measurement data from a three-lane unidirectional motorway, M40, outside Madrid, Spain from 8:30 a.m. to 9:00 a.m. (busy hour), and from 11:30 a.m. to 12:00 p.m. (off-peak) on May 7 2010. Sensors buried under the concrete layer of the roadway collected per-lane measurements every second about the number of passing vehicles and their speed. The measurements calibrated a microscopic simulator, whose output is the location of each vehicle, i.e., lane and horizontal position over a road segment of $10$ km with one second granularity. The corresponding simulation time is half an hour, which means that $1\,800$ snapshots of vehicles per time span are available. The traces have been calibrated to represent quasi-stationary road traffic for each lane~\cite{Fiore2014}. We will drop $600$ snapshots from the initial ten minutes of the simulation, in order to allow the first vehicle entering the roadway reach at the exit. This is to have enough samples of inter-vehicle distances for each snapshot while constructing its empirical \ac{CDF}. We will fit the \ac{PPP} and the hardcore point process for each snapshot and lane, independently of other snapshots and lanes. The interference and outage models in the following sections are for a single snapshot too. Modeling the temporal aspects of interference, see for instance~\cite{Gong2014, Koufos2016}, using the vehicle trajectory is left for future work.
\begin{figure*}[!t]
 \centering \subfloat[Left lane]{\includegraphics[width=2in]{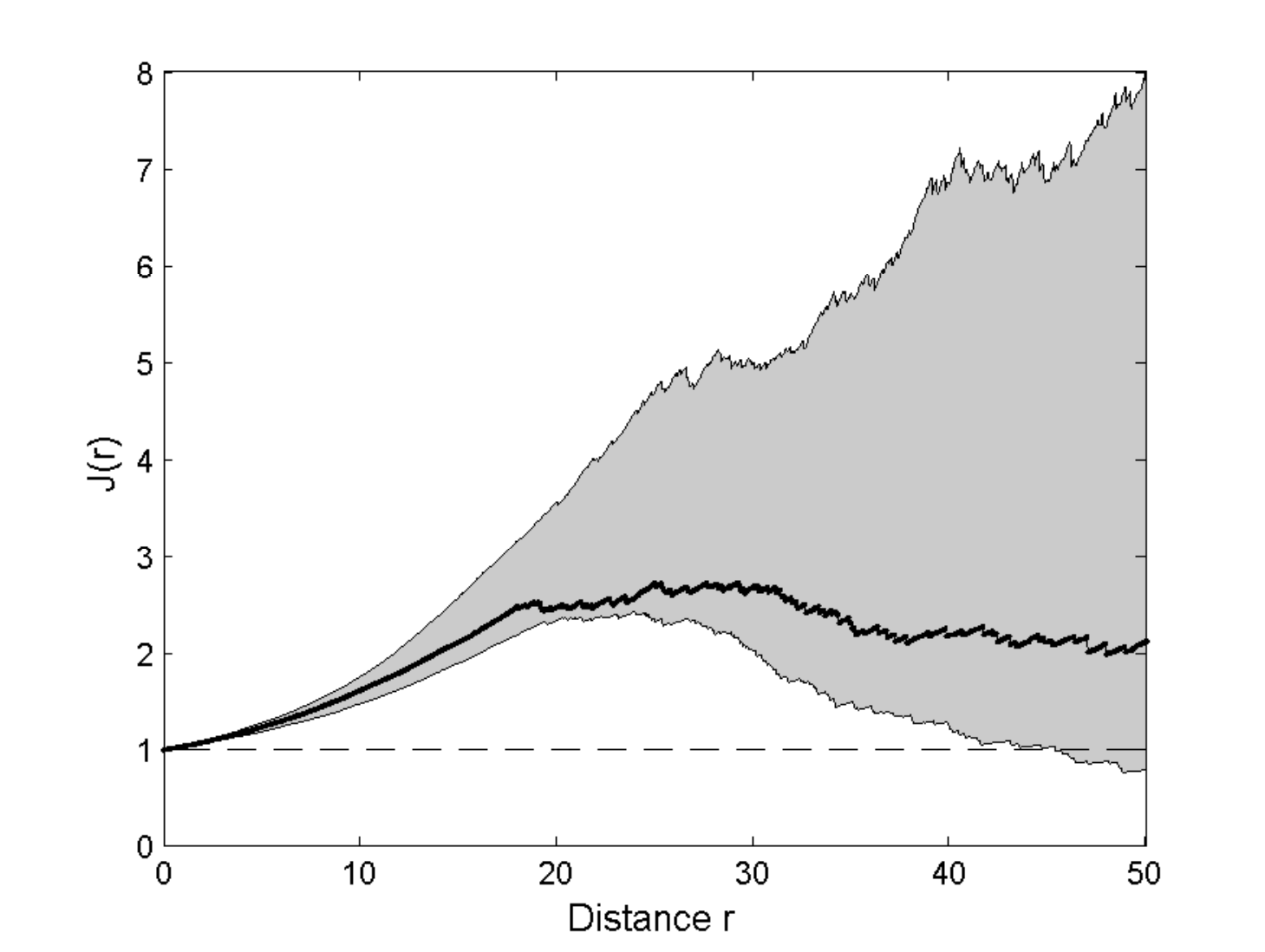}\label{fig:JrLeft}}\hfil \subfloat[Middle lane]{\includegraphics[width=2in]{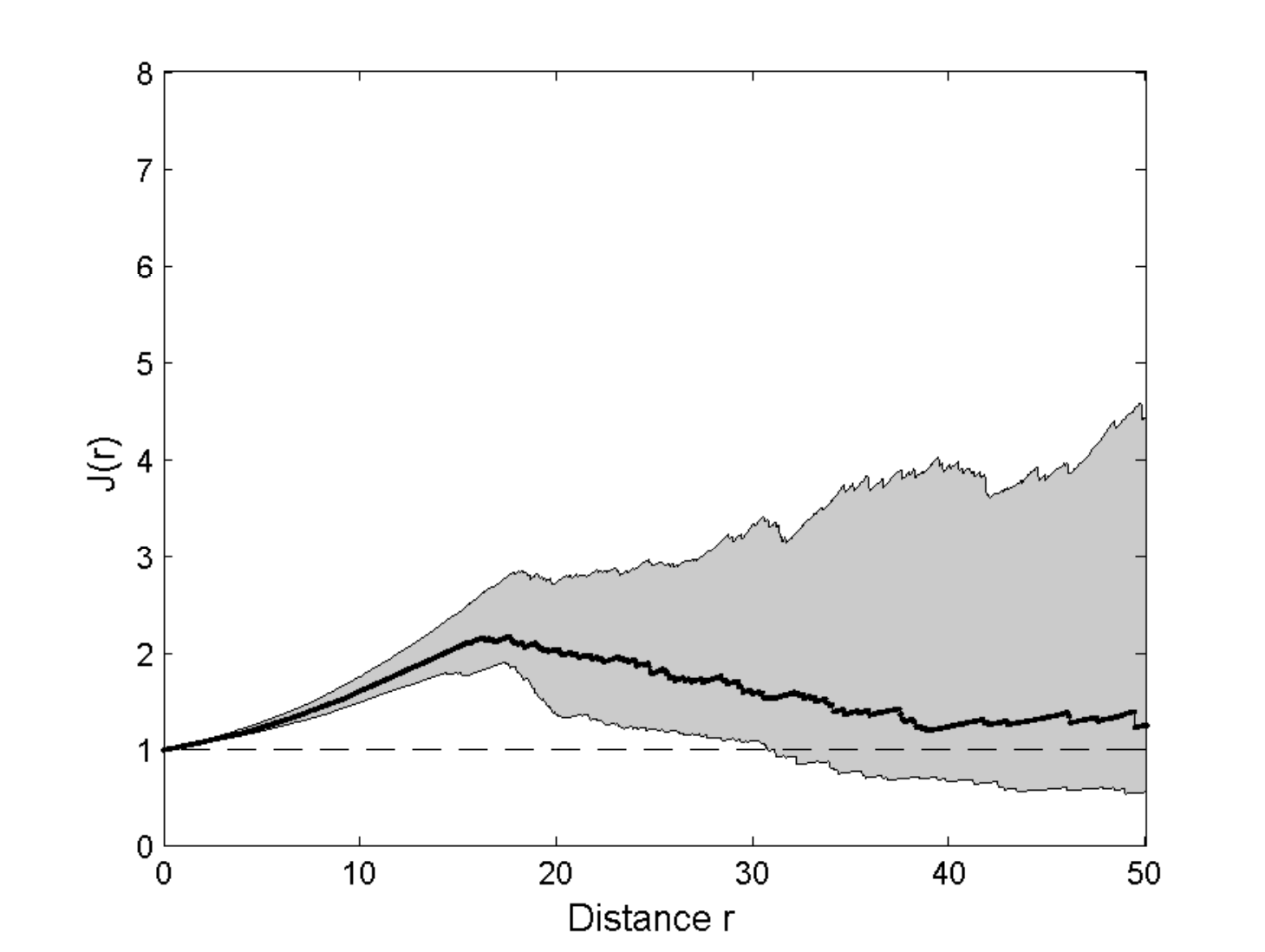}}\hfil \subfloat[Right lane]{\includegraphics[width=2in]{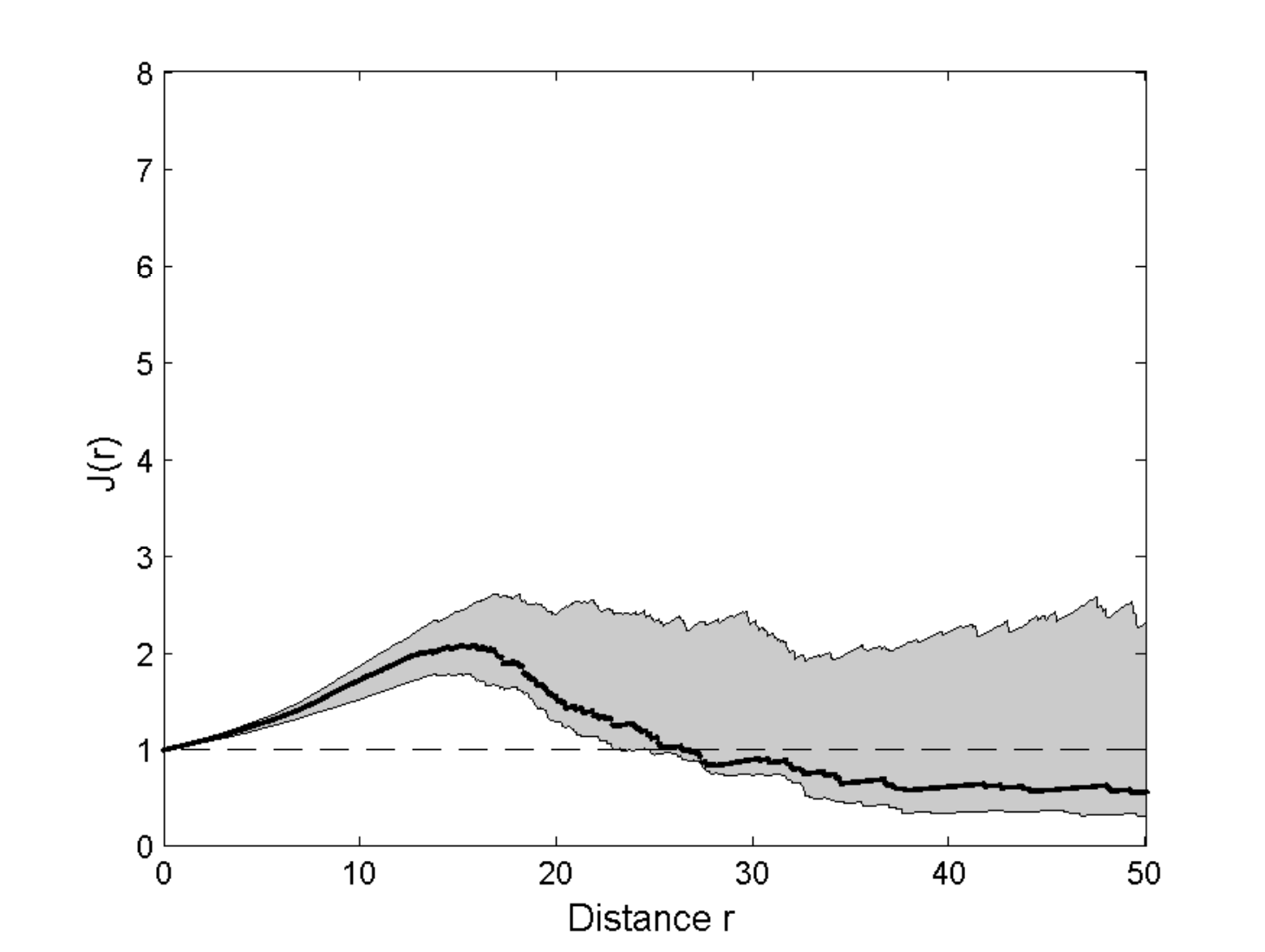}} 
 \caption{Envelope of J-function over 1200 snapshots for the busy hour (gray-shaded areas). The solid line is an example illustration of ${\text{J}}\!\left(r\right)$ for the 1000-th snapshot. The \acp{CDF} $G,F$ are estimated within a window of $9$ km to avoid boundary effects. The \ac{PPP} corresponds to ${\text{J}}\!\left(r\right)\!=\!1$, dashed-line.}
 \label{fig:Jr}
\end{figure*}
\begin{figure*}[!t]
 \centering \subfloat[Left lane]{\includegraphics[width=2in]{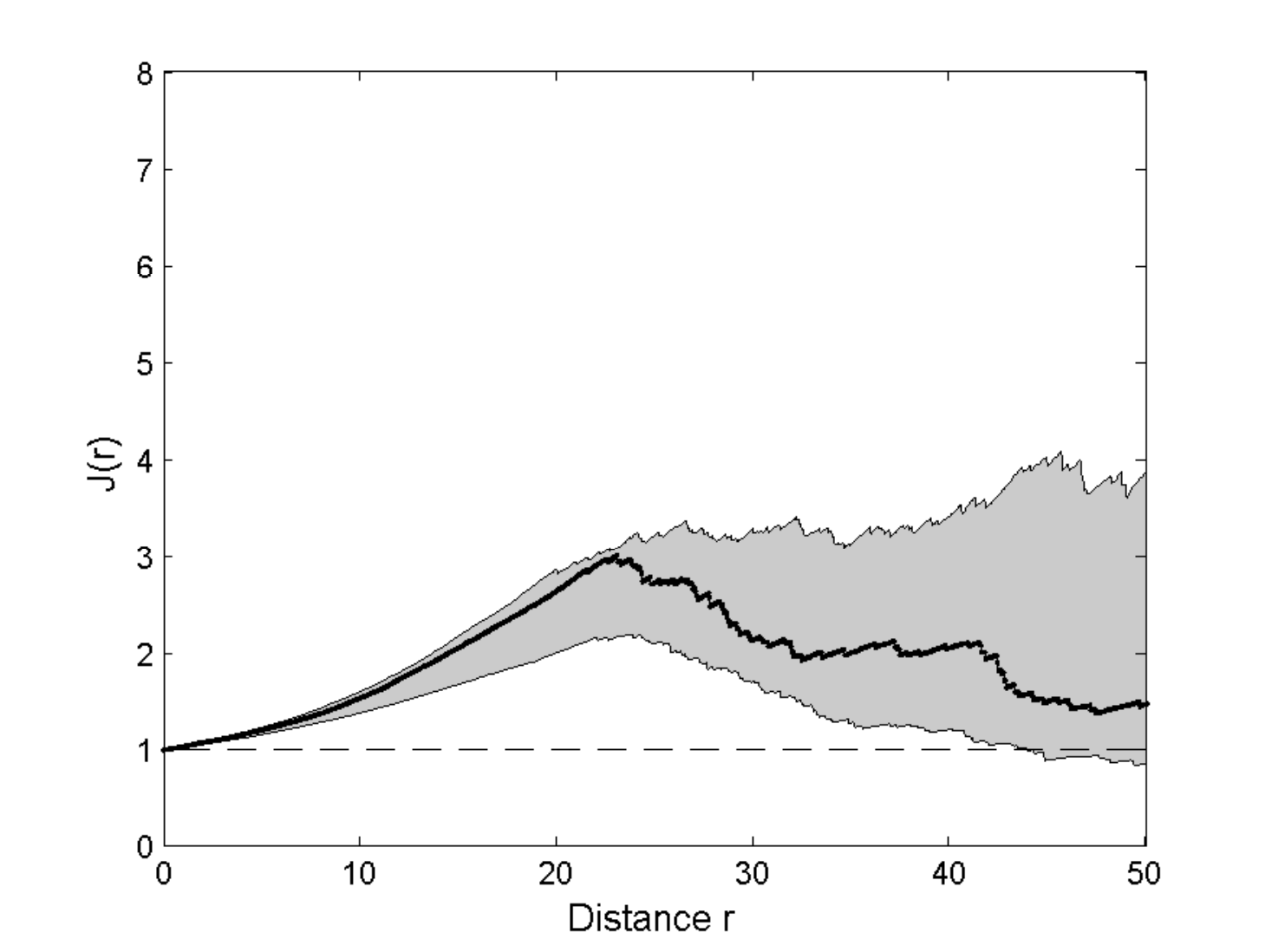}}\hfil \subfloat[Middle lane]{\includegraphics[width=2in]{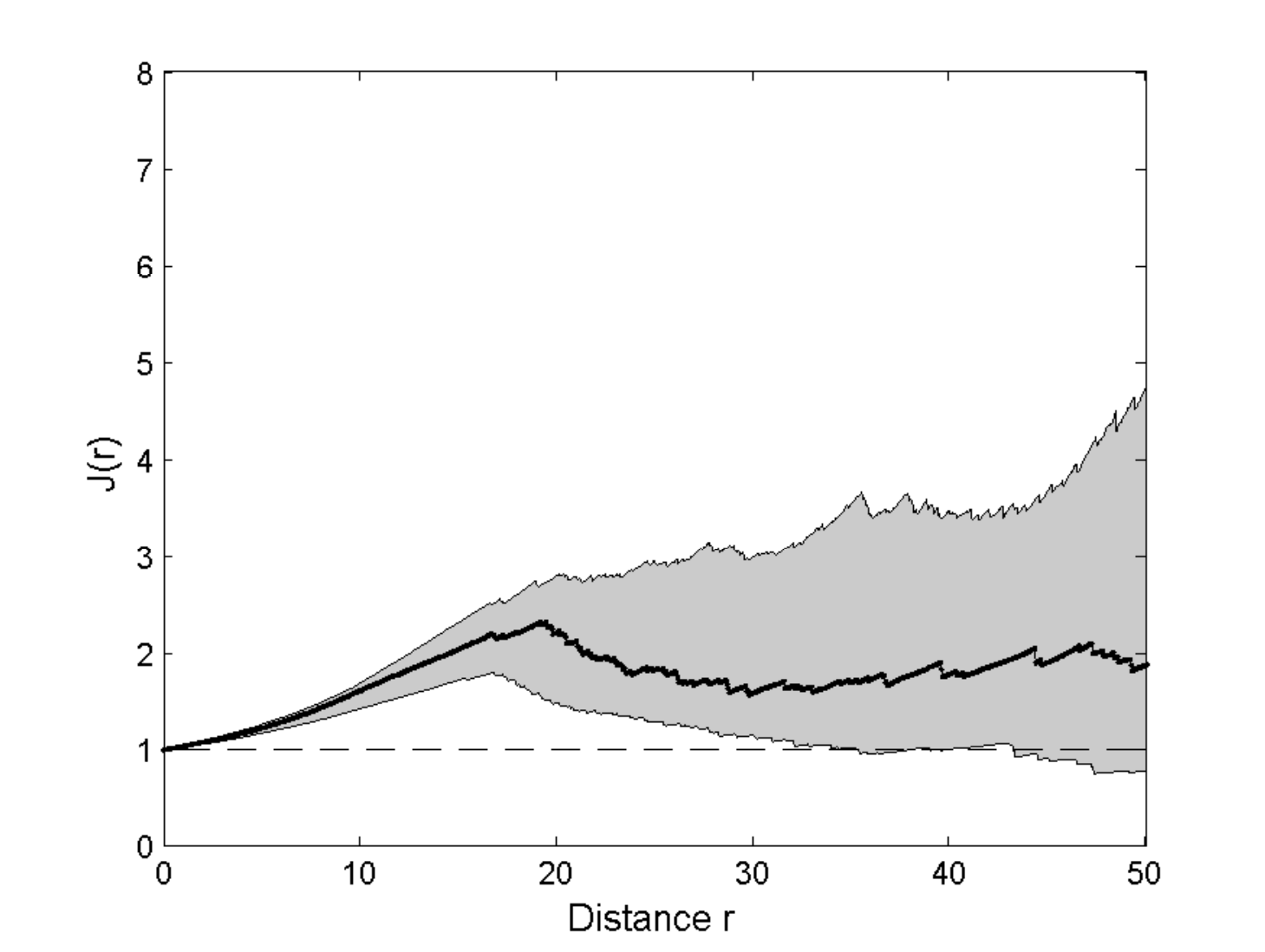}}\hfil \subfloat[Right lane]{\includegraphics[width=2in]{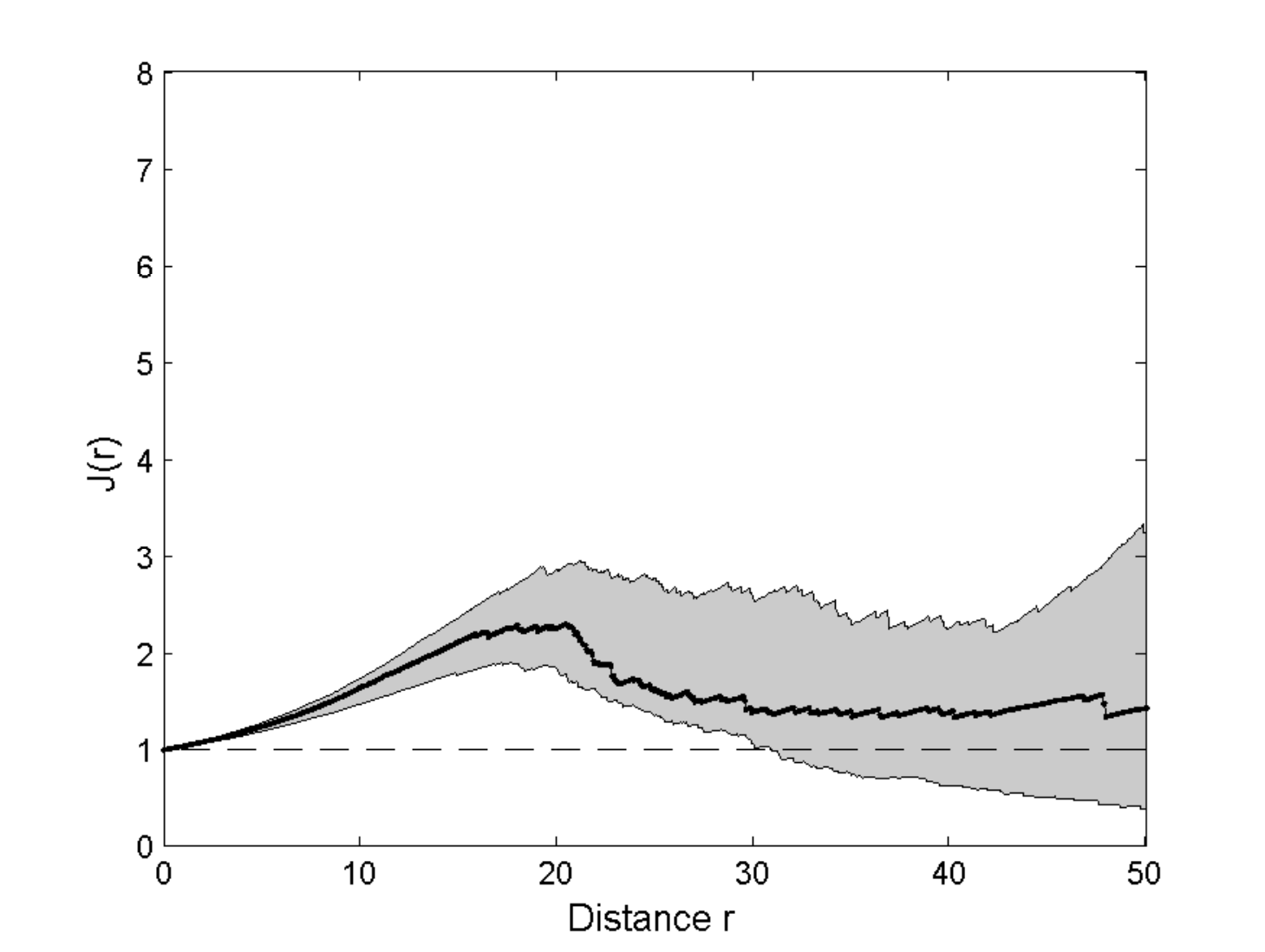}} 
 \caption{Envelope of J-function over 1200 snapshots during off-peak. See the caption of previous figure for explanation of the lines.}
 \label{fig:JrOff}
\end{figure*}

\begin{figure*}[!t]
 \centering \subfloat[Left lane]{\includegraphics[width=2in]{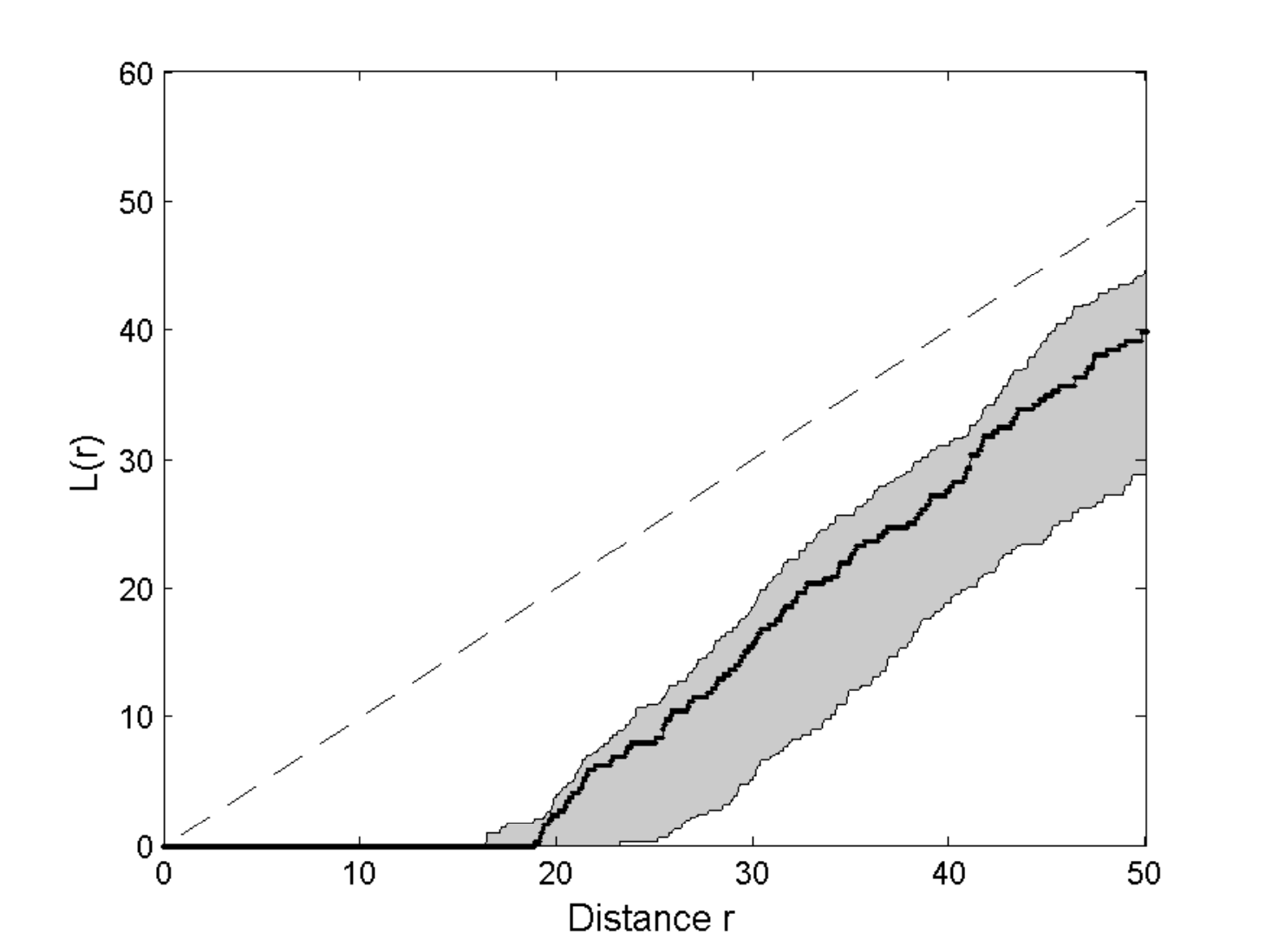}\label{fig:KtLeft}}\hfil \subfloat[Middle lane]{\includegraphics[width=2in]{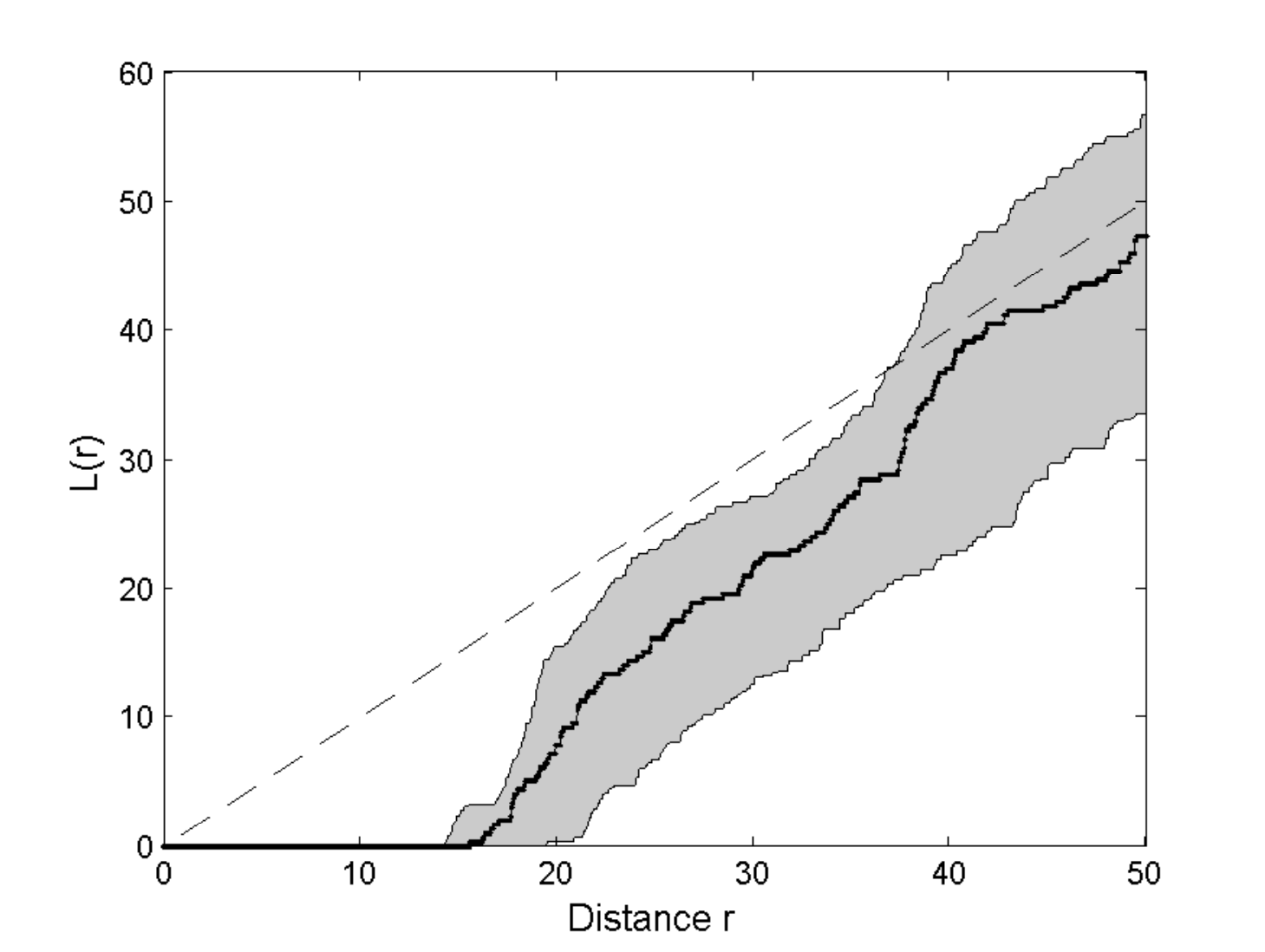}}\hfil \subfloat[Right lane]{\includegraphics[width=2in]{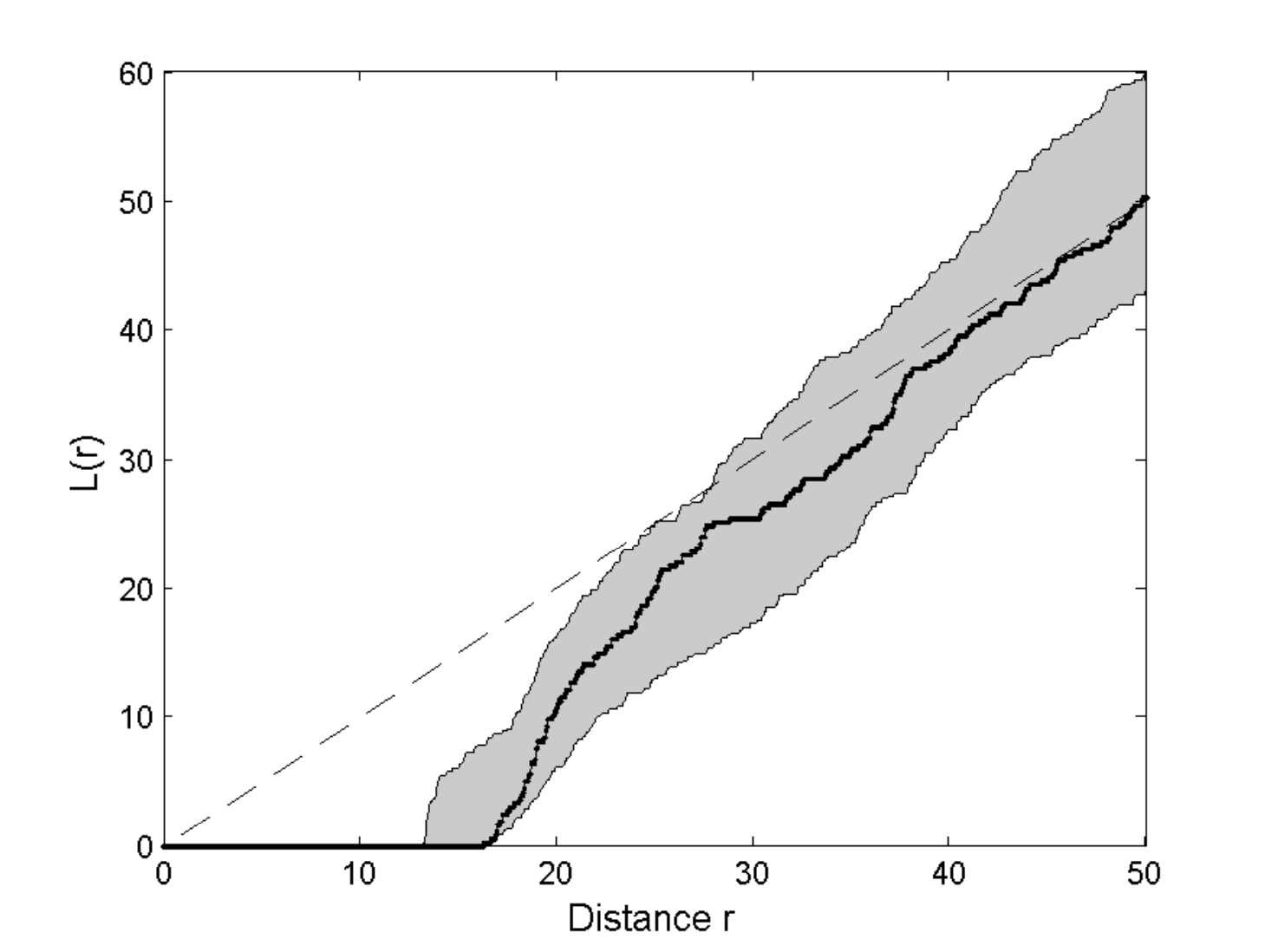}} 
 \caption{Envelope of ${\text{L}}\!\left(r\right)\!=\!\frac{{\text{K}}\!\left(r\right)}{2}$ over 1200 snapshots for the busy hour. Example illustration of ${\text{L}}\!\left(r\right)$ for the 1000-th snapshot (solid line). For each snapshot the intensity per lane required in the calculation of the K-function is estimated as being the inverse of the mean inter-vehicle distance. The L-function is evaluated within an inner window of $9$ km to avoid boundary effects, $r_{\text{max}}\!=\! 500$ m. The dashed-line corresponds to \ac{PPP}.}
 \label{fig:Kt}
\end{figure*}

In order to simulate the per-lane distributions $G_i, F_i, i\!\in\!\left\{1,2,3\right\}$ for a snapshot, we first take an inner segment (or window) of the $i$-th lane to avoid boundary effects. The endpoints of the window must be at least $r_{\text{max}}$ away from the endpoints of the roadway, where $r_{\text{max}}$ is the maximum considered $r$ in the evaluation of the J-function~\cite[Chapter 1.10]{Baddeley2007}. In order to simulate the distribution of $G_i$, we calculate the nearest-neighbor distance for each point in the window. In order to simulate the distribution of $F_i$, we randomly distribute $10^4$ locations within the window, and for each of them we calculate its minimum distance over the snapshot's points. In Fig.~\ref{fig:Jr}, it is evident that all lanes exhibit repulsion for small distances, ${\text{J}}\!\left(r\right)\!>\!1, r\!\leq\! 20 {\text{m}}$. The repulsion is stronger at the left lane, because this lane experiences the highest average speeds~\cite[Fig.~5b]{Fiore2015}: The envelope stays larger than unity up to $40$ m and there are only few instances of clustering for the considered range of distances. The J-function for the right lane indicates that the traces may also exhibit clustering for distance $r\!>\! 20$ m. Nevertheless, for small distances, the empirical J-function behaves similarly to the J-function of the hardcore process, see Fig.~\ref{fig:JfunctionHC}. This is due to the safety distance the drivers maintain from the vehicle ahead. During off-peak, there is still clear repulsion at small distances, but the differences between the lanes are less prominent. Off-peak is characterized by lower intensity of vehicles. Because of that, the drivers select their lane more freely as compared to the busy hour, making the lanes to look more alike to each other. In Fig.~\ref{fig:Kt} we reconfirm, through the envelope of L-function, that the left lane has the highest repulsion. In order to simulate the L-function for a snapshot, we take each point in the window and count the number of points within a distance $r$ from that point. Then, we average over all points and normalize by $2\lambda$. The envelope quantifies the range of hardcore distance for each lane. As expected, the left lane exhibits the highest values.

The summary statistics have justified that a hardcore process is more suitable to model the available traces than a \ac{PPP}. We will use Cowan M2 for the locations of vehicles along each lane. Apart from being used in transportation research~\cite{Cowan1975}, this model will allow us to construct simple approximations for the moments of interference in the following sections. Before starting with interference modeling, we still need to estimate the model parameters for a snapshot, i.e, per-lane intensity and hardcore distance. Also, it is worth demonstrating whether the summary statistics for a snapshot fall within the simulated envelope of the fitted hardcore process.
\begin{figure*}[!t]
 \centering \subfloat[Left lane]{\includegraphics[width=2in]{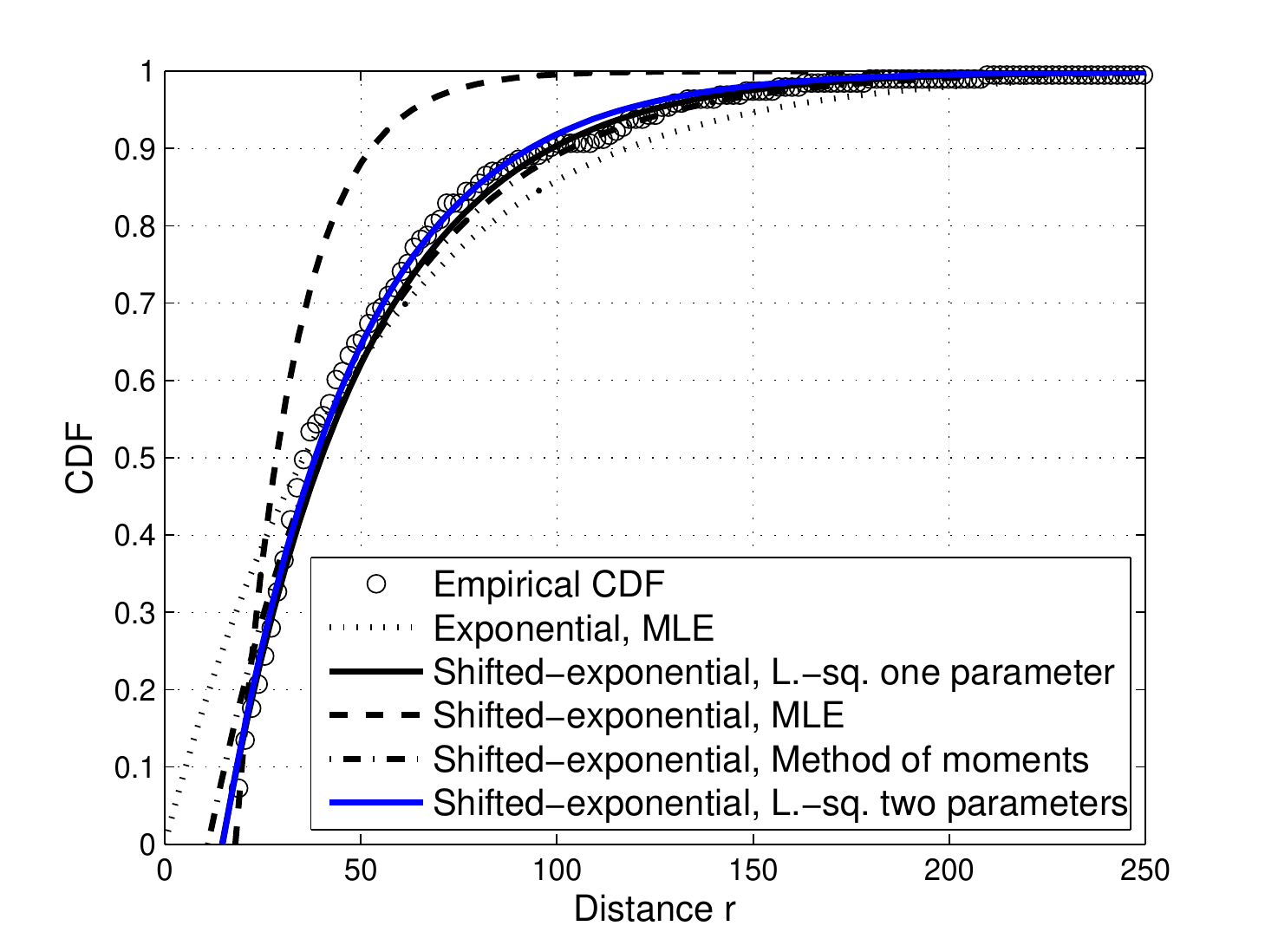}}\hfil \subfloat[Middle  lane]{\includegraphics[width=2in]{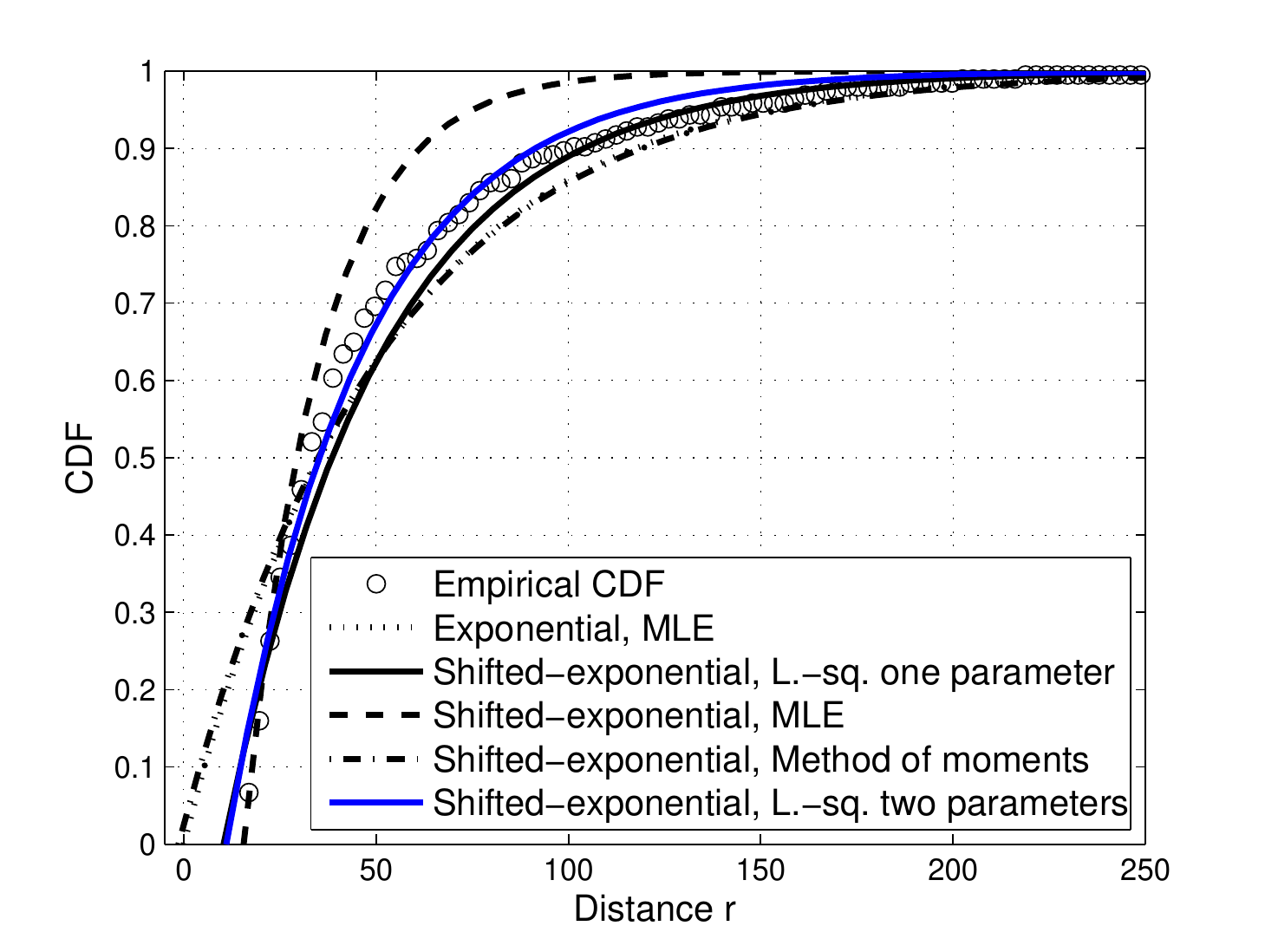}\label{fig:MiddleLane}}\hfil \subfloat[Right lane]{\includegraphics[width=2in]{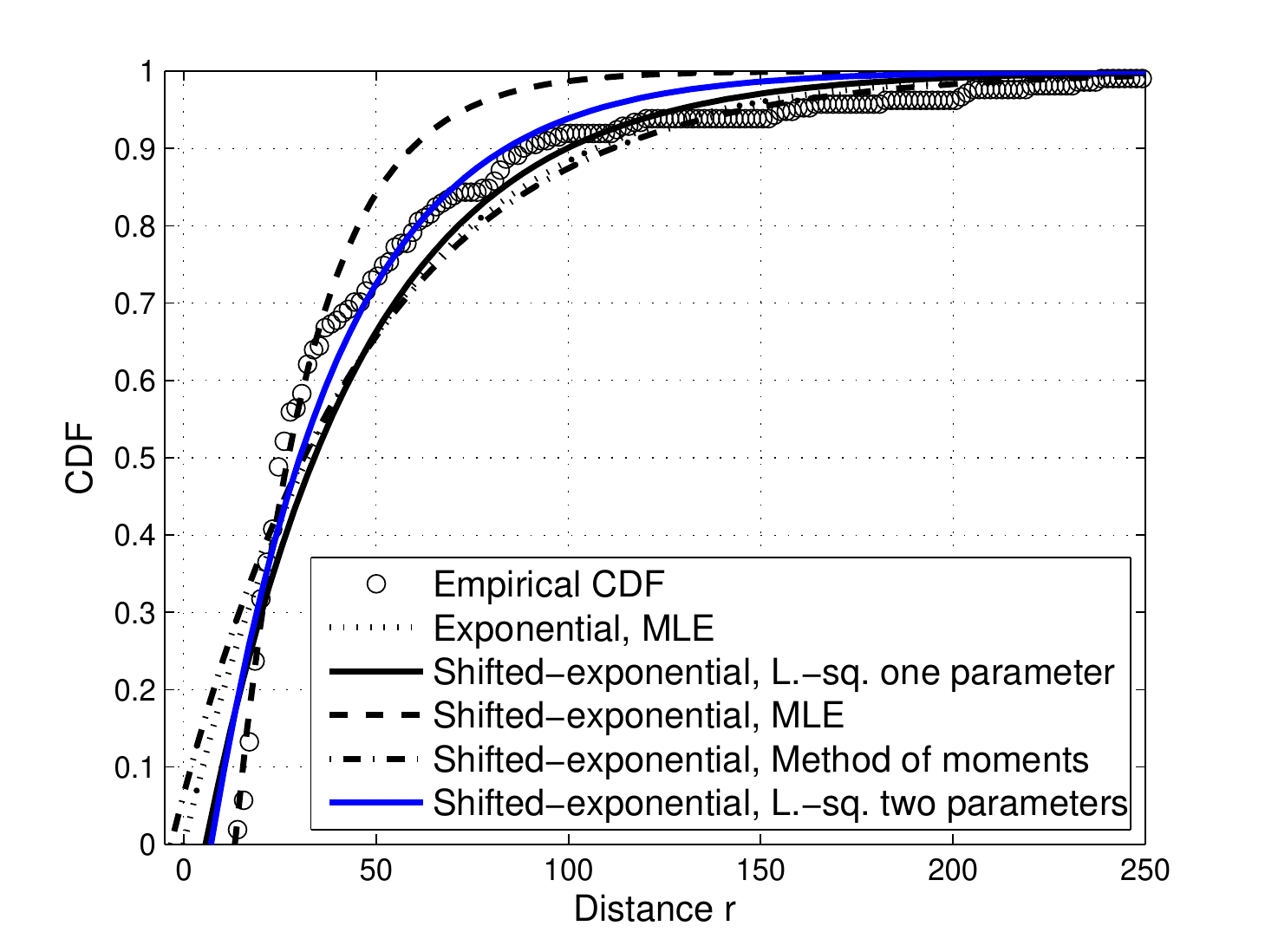}\label{fig:RightLane}} 
 \caption{Empirical \ac{CDF} of inter-vehicle distances at the 1000-th snapshot of the busy hour along with approximations.}
 \label{fig:ThreeLanesDistApprox}
\end{figure*}
\begin{figure*}[!t]
 \centering \subfloat[Intensity $\lambda_i$]{\includegraphics[width=2in]{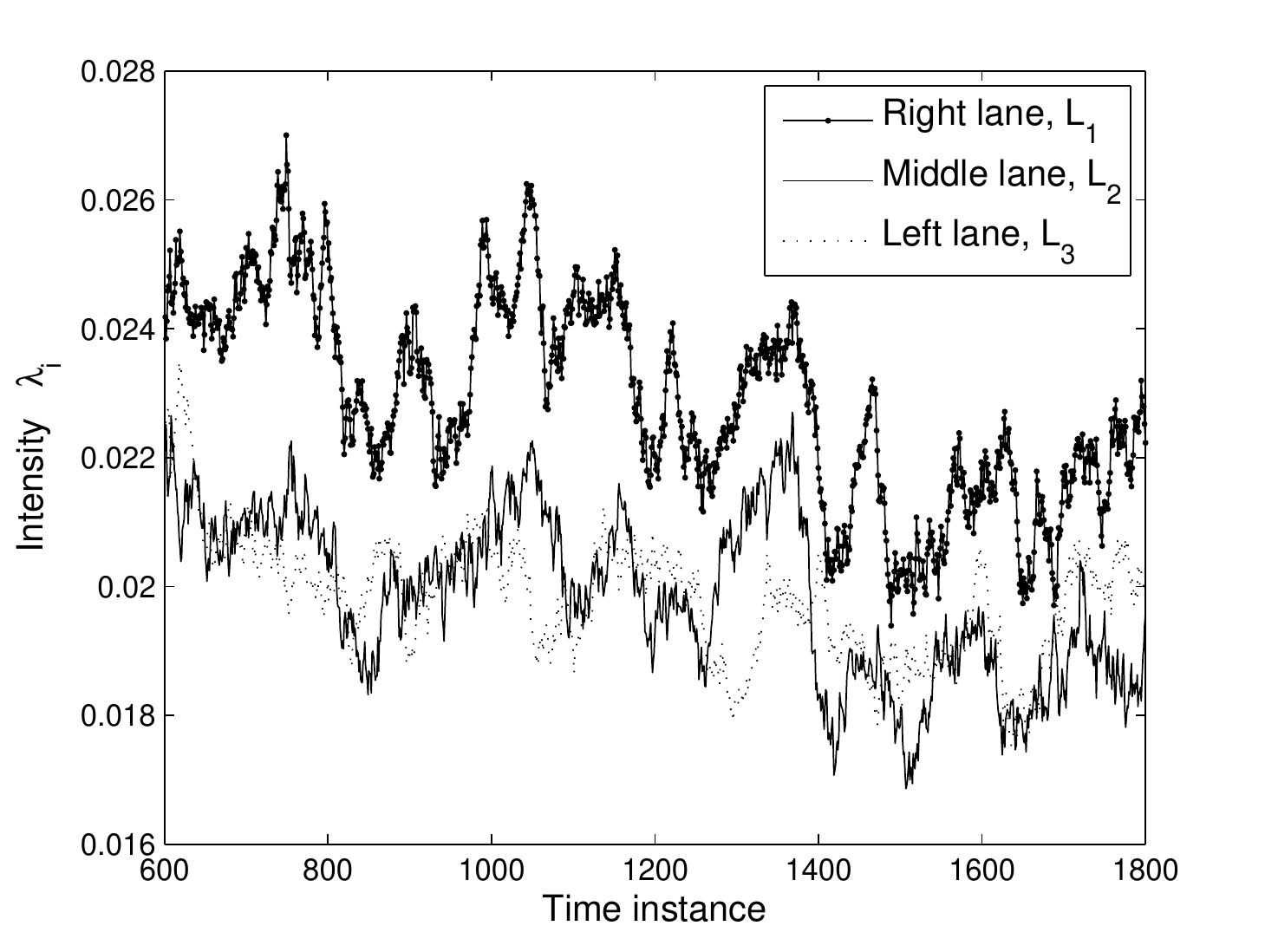}}\hfil \subfloat[Hardcore $c_i$]{\includegraphics[width=2in]{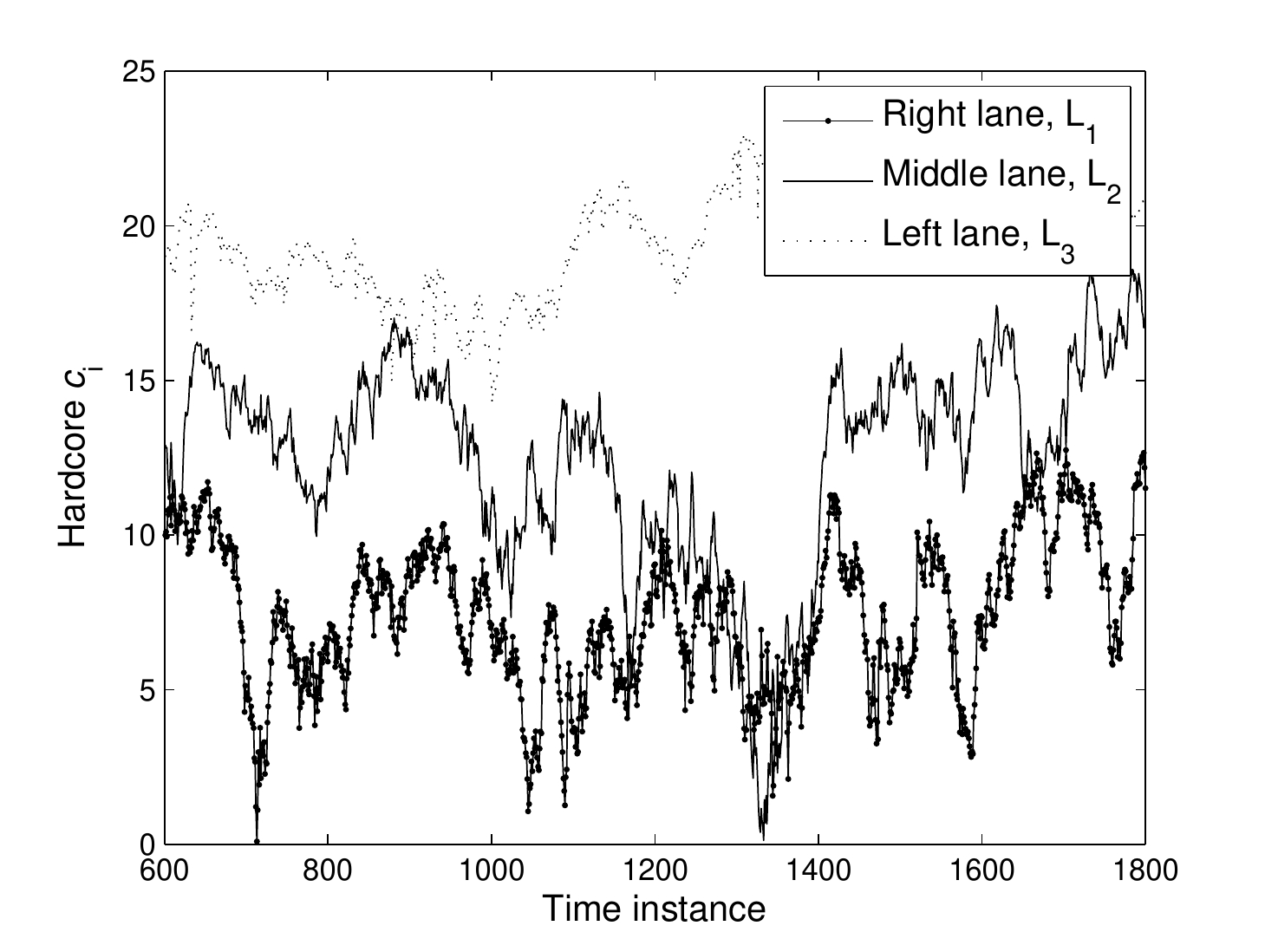}}\hfil \subfloat[Product $\lambda_i c_i$]{\includegraphics[width=2in]{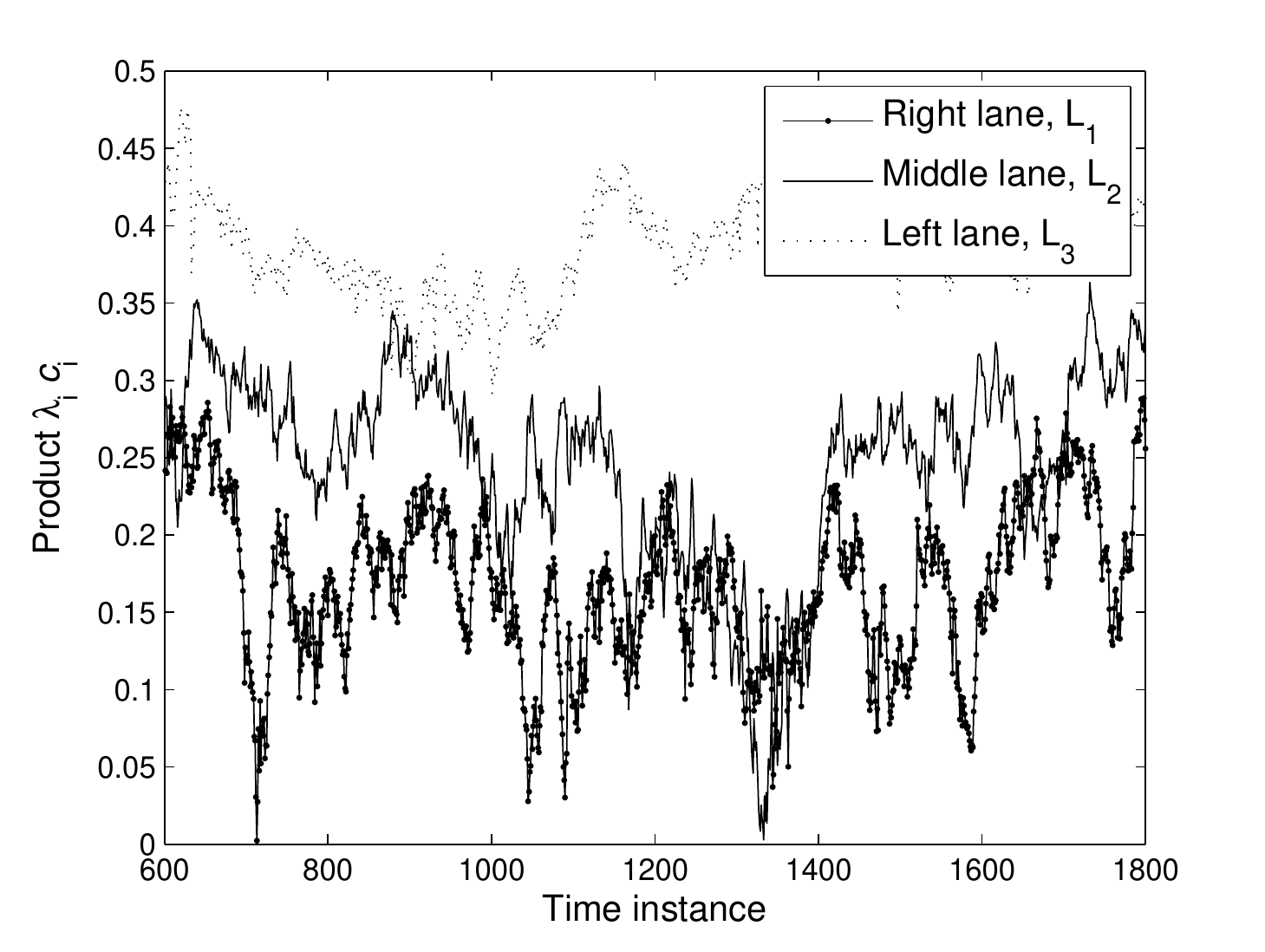}} 
 \caption{Estimation of intensity and hardcore distance per lane using synthetic traces between 8:40 a.m. and 9:00 a.m.}
 \label{fig:Parameter}
\end{figure*}
\begin{figure*}[!t]
 \centering \subfloat[Intensity $\lambda_i$]{\includegraphics[width=2in]{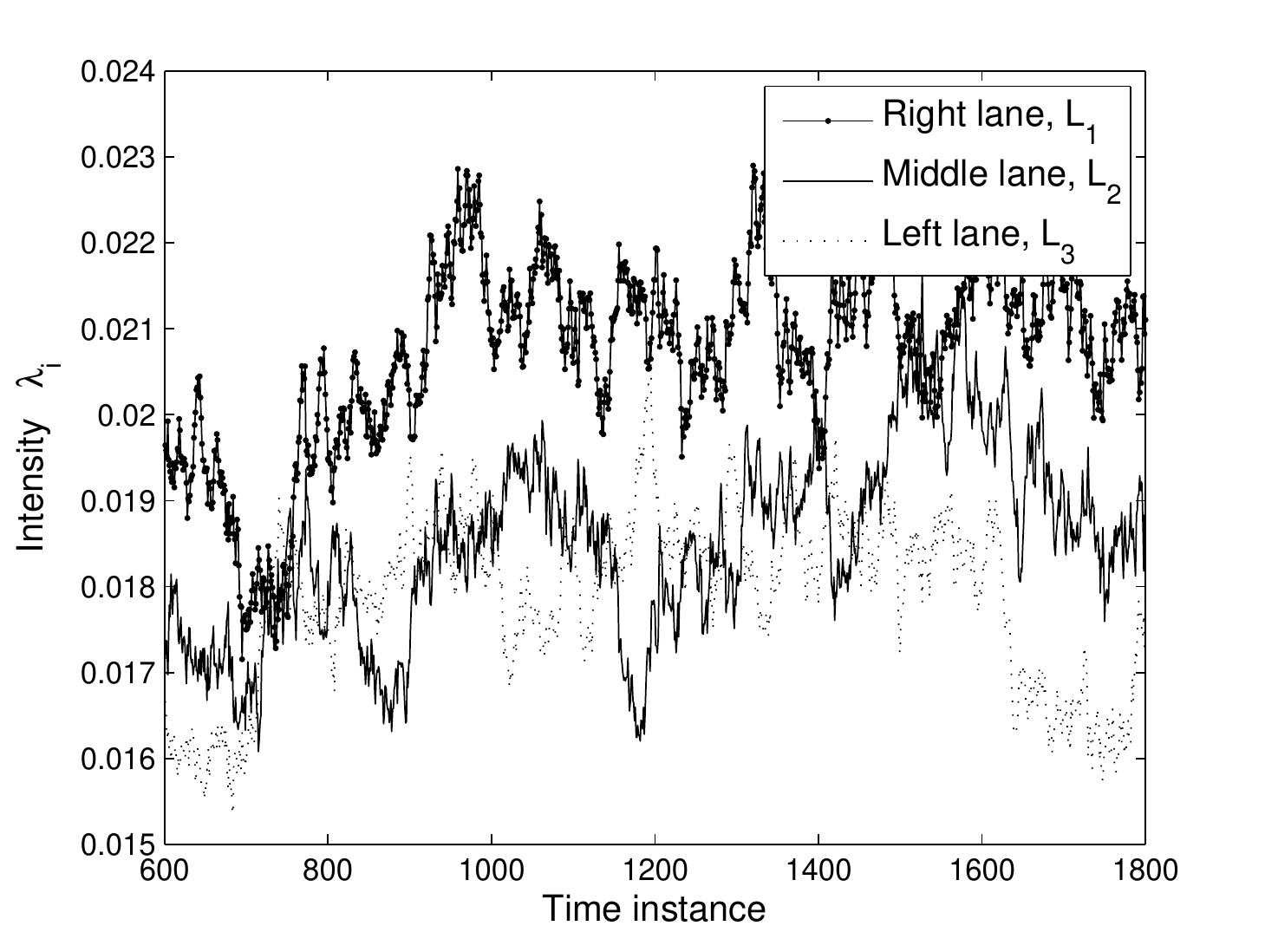}}\hfil \subfloat[Hardcore $c_i$]{\includegraphics[width=2in]{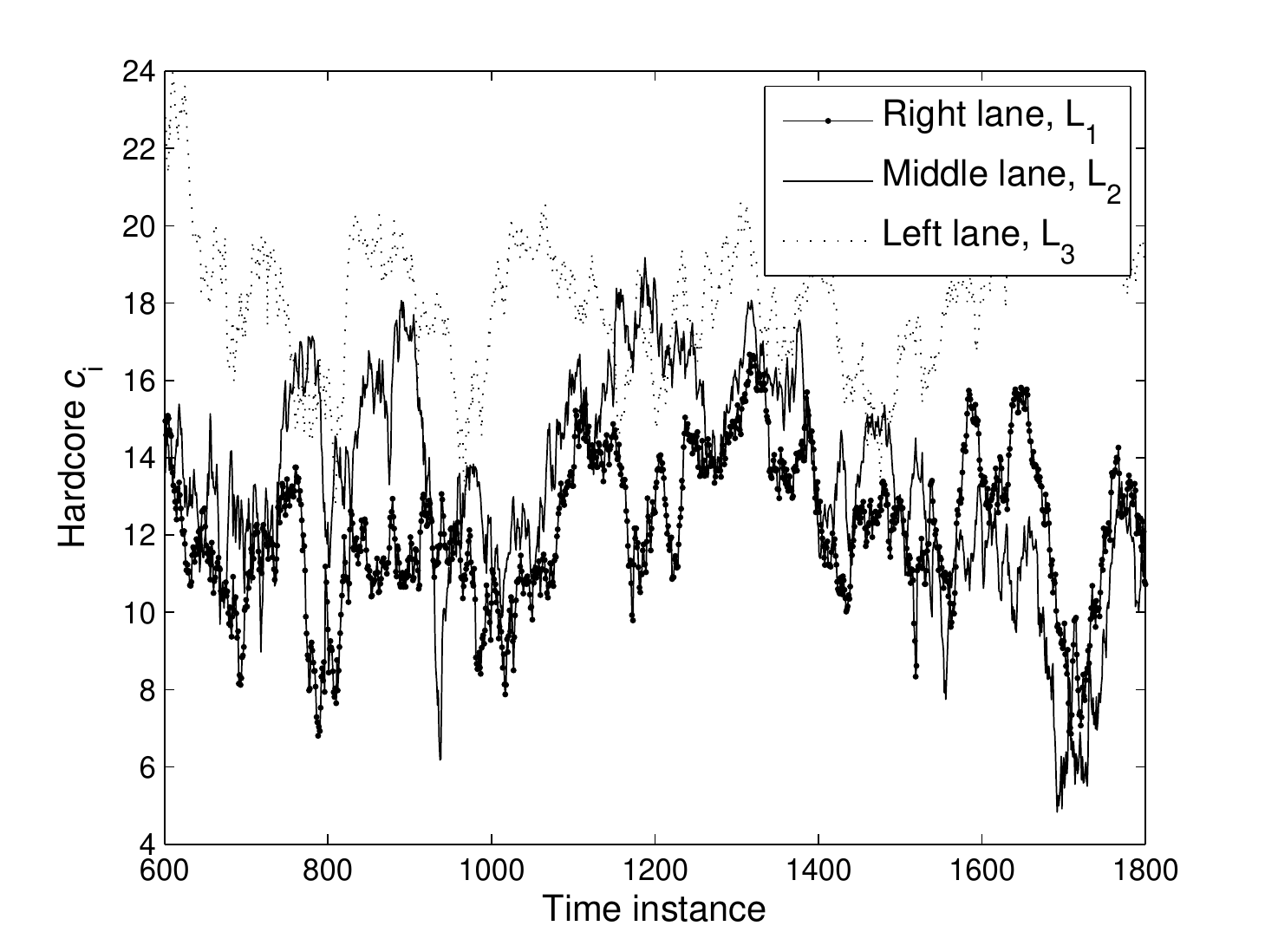}}\hfil \subfloat[Product $\lambda_i c_i$]{\includegraphics[width=2in]{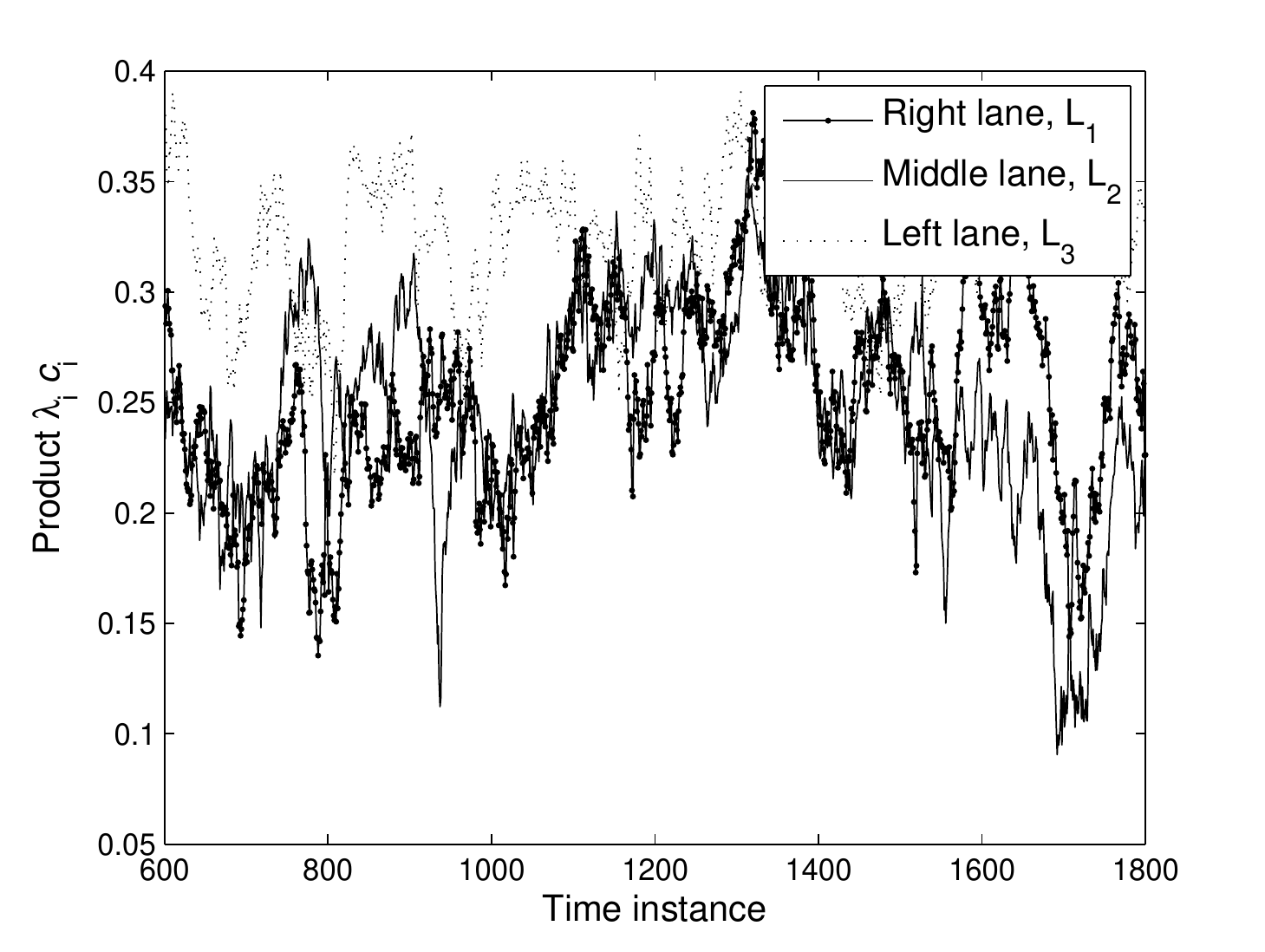}} 
 \caption{Estimation of intensity and hardcore distance per lane using synthetic traces between 11:40 a.m. and 12:00 p.m.}
 \label{fig:Parameter2}
\end{figure*}

Let us denote the inter-vehicle distances for the $i$-th lane $i\!\in\!\left\{1,2,3\right\}$ by $z_{i,j}, j\!=\!1,2,\ldots n_i$, where $n_i$ is the sample size. For the \ac{PPP}, we will estimate the intensity as being equal to the inverse of the mean inter-vehicle distance obtained from the sample, i.e., $\hat{\lambda}_i\!=\!\left(\frac{1}{n_i} \sum_{j=1}^{n_i}z_{i,j}\right)^{-1}$, which is the \ac{MLE}. For the hardcore process, we will parameterize the intensity and the hardcore distance using various methods. (i) The \ac{MoM} matches the mean, $\left(c_i\!+\!\mu_i^{-1}\right)$, and the variance, $\mu_i^{-2}$, of the shifted-exponential distribution to the sample mean and variance. (ii) The \ac{MLE} is the minimum inter-vehicle distance obtained from the sample $\hat{c}_i\!=\!\min_j\left\{z_{i,j}\right\}$, and $\hat{\lambda_i}\!=\!\left(\frac{1}{n_i}\sum_{j=1}^{n_i}z_{i,j}\!-\!\hat{c}_i \right)^{-1}$. (iii) The (non-linear) least-squares iteratively estimate the $\hat{\mu}_i,\hat{c}_i$ minimizing the square difference between the empirical \ac{CDF} and the shifted-exponential \ac{CDF}, $\left(1\!-\!e^{-\hat{\mu}_i\left(x_d-\hat{c}_i\right)}\right)$, where $x_d$ are the bins. In order to reduce computational complexity, we may set first $\hat{\lambda}_i\!=\!\left( \frac{1}{n_i}\sum_{j=1}^{n_i}z_{i,j}\right)^{-1}$, then estimate a single parameter $c_i$. In that case, the \ac{PPP} and the hardcore process are forced to have the same intensity. In either case, we must constrain $0\!\leq\!\hat{c}_i\!\leq\!\hat{\lambda}_i^{-1}$. 

In Fig.~\ref{fig:ThreeLanesDistApprox} we have plotted the empirical \ac{CDF} of inter-vehicle distances for a snapshot. The accuracy of the various fitting methods is similar for other snapshots too. The exponential \ac{CDF} cannot capture at all the repulsion between successive vehicles. The \ac{MoM} may give a negative estimate for the hardcore distance, see Fig.~\ref{fig:MiddleLane} and Fig.~\ref{fig:RightLane}. The \ac{MLE} for the shifted-exponential distribution fits very well the lower tail but it fails elsewhere. The least-squares estimation provides relatively good fit over the full range. When $\hat{\lambda}_i$ is fixed equal to the \ac{MLE} of \ac{PPP}, the fit becomes slightly worse. 

In Fig.~\ref{fig:Parameter}, we depict the estimates for the intensity and the hardcore distance over $1\,200$ snapshots using  least-squares, implemented in the curve fitting toolbox in MatLab. The right lane has the highest intensity and the left lane, due to the high speeds, gives the highest values for the hardcore distance and for the product $\lambda c$. During off-peak, see Fig.~\ref{fig:Parameter2}, the discrepancy of $\hat{\lambda}_i,\hat{c}_i$ between the lanes becomes less prominent. This is in accordance with the behavior of the J-function during busy hour and off-peak, see Fig.~\ref{fig:Jr} and Fig.~\ref{fig:JrOff}. 

\begin{figure*}[!t]
 \centering \subfloat[Left lane]{\includegraphics[width=2in]{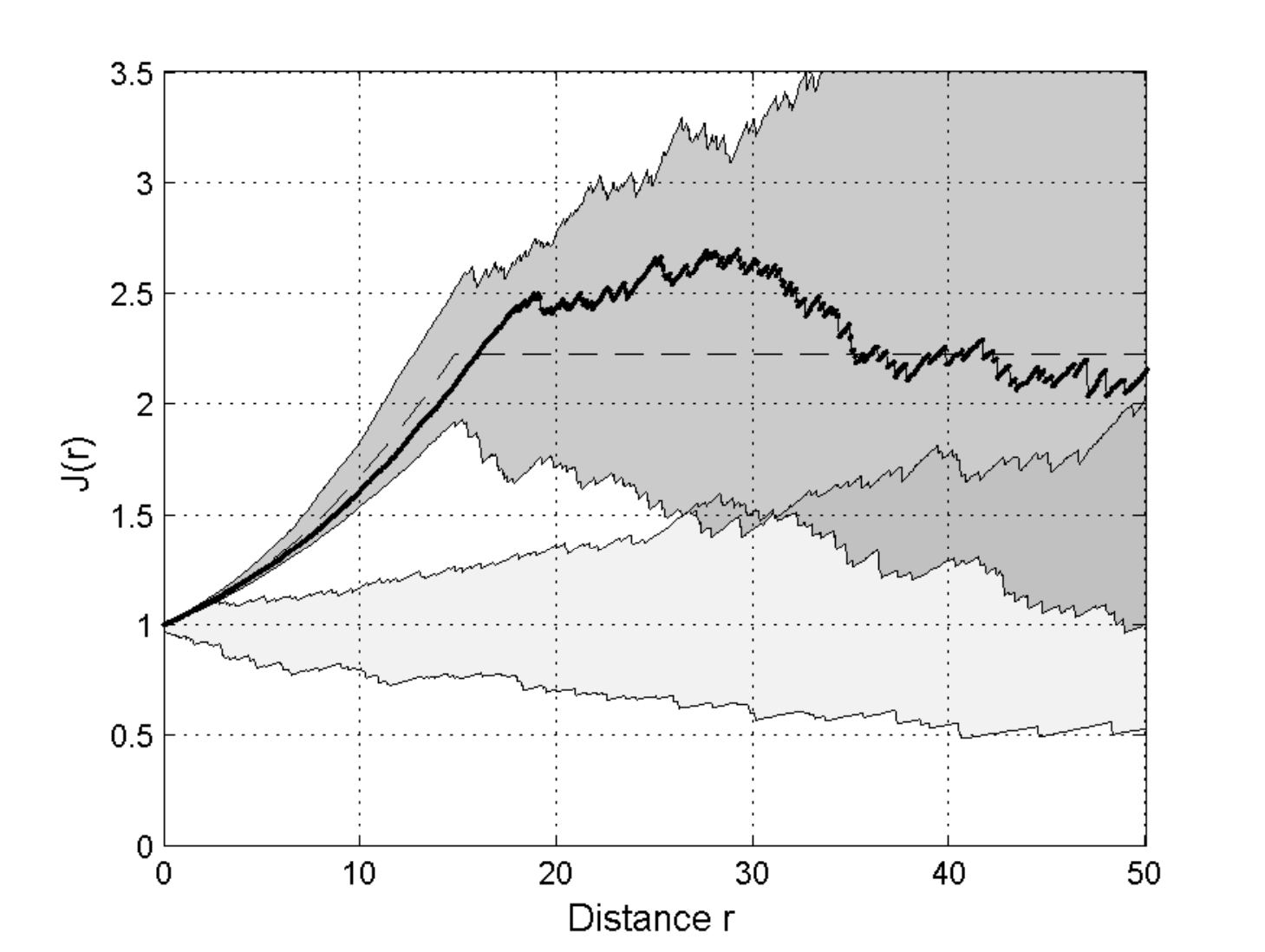}}\hfil \subfloat[Middle lane]{\includegraphics[width=2in]{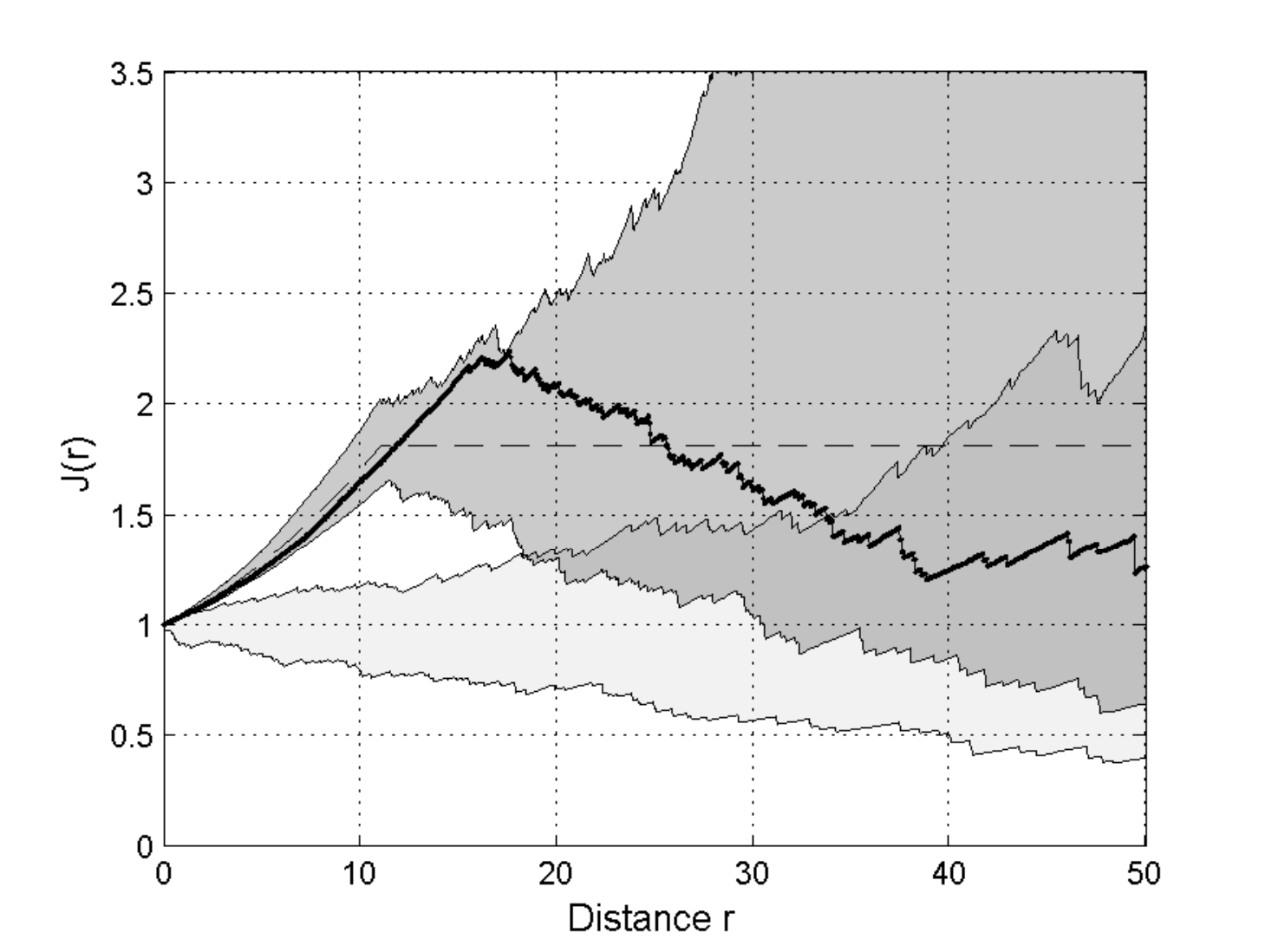}}\hfil \subfloat[Right lane]{\includegraphics[width=2in]{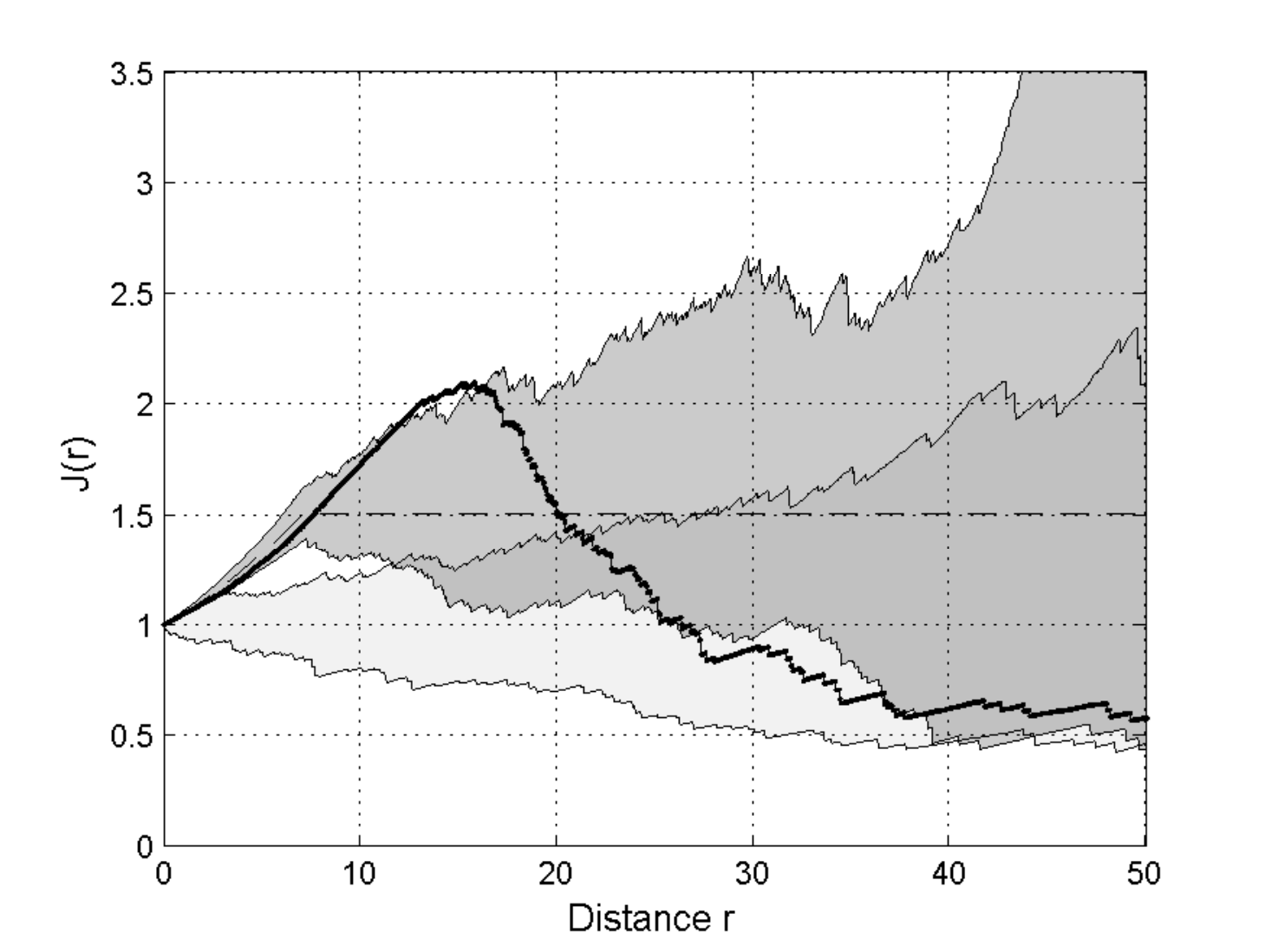}} 
 \caption{The empirical J-function for the 1000-th snapshot of the busy hour (solid line), the calculated ${\text{J}}\!\left(r\right)$, see equation~\eqref{eq:Jfunction}, for the fitted hardcore point process (dashed-line), the simulated envelope of the J-function for the fitted hardcore process over $99$ runs (gray-shaded area), and the simulated envelope of the J-function for the fitted \ac{PPP} over $99$ runs (lightly gray-shaded area).}
 \label{fig:SimHardcoreJ}
\end{figure*}

Finally, in Fig.~\ref{fig:SimHardcoreJ}, we illustrate that the envelope of J-function for the fitted hardcore process contains the J-function generated by the snapshot. Only for the right lane, the snapshot may fall slightly outside the envelope. On the other hand, the envelope of J-function using the fitted \ac{PPP} cannot capture at all the J-function of the snapshot for all lanes and distances $r\!\leq\! 20$ m in Fig.~\ref{fig:SimHardcoreJ}. This is another evidence about the suitability of the hardcore process to model motorway traffic in comparison with \ac{PPP}. Even though the calculated ${\text{J}}\!\left(r\right)$ for the fitted hardcore process, see~\eqref{eq:Jfunction}, matches that of the snapshot only in the initial increasing part, we will illustrate in the next section that Cowan M2 considerably improves the outage probability predictions of \ac{PPP}. 

\section{Outage probability under reduced Palm}
\label{sec:MeanVariance}
The \ac{PGFL} of the hardcore point process which can be used to calculate the \ac{LT} of interference is not available. Also, due to the complicated form of the \ac{PCF}, only the first few terms of the factorial moment expansion of the \ac{PGFL}, see~\cite{Westcott1972}, can be approximated. The simplest way to get around these issues, is to calculate a few moments and fit the interference distribution to well-known functions with simple \acp{LT}. The \ac{MoM} has been widely-used in wireless communications research to model signal-to-noise ratio in composite fading channels~\cite{Atapattu2011}, aggregate interference~\cite{Sousa2008} and spectrum sensing~\cite{Koufos2011}. 

In order to select appropriate candidate distributions, we first note that the interference \ac{CDF} decays exponentially fast near the origin because the point process is stationary~\cite[Theorem 4]{Ganti2008}. In addition, for bounded propagation pathloss model (the hardcore distance $c$ essentially makes the pathloss function bounded), the decay at the tail is dominated by the fading distribution~\cite[Theorem 3]{Ganti2008}, thus this is exponential too. Popular distributions for interference modeling in wireless networks with irregular geometry can be found in~\cite{Haenggi2015, Kountouris2014} including gamma, inverse Gaussian and Weibull. Even though the inverse Gaussian distribution seems the best candidate because it decays exponentially fast near the origin and at the tail, it has only two parameters, and it does not provide very good fit via \ac{MoM} in our system. Its generalized counterpart has three parameters, but numerical methods are required to calculate them. Instead, we will use the shifted-gamma distribution which allows us to express its parameters in a simple form. The gamma distribution decays polynomially at the origin and thus, we expect to see some discrepancy in the upper tail of the \ac{SIR} \ac{CDF}. In some recent work~\cite{Vu2017}, the parameters of the approximating distribution have been calculated using a combination of moment matching and \ac{MLE}. An apparent advantage of this method is its extention to mixture models using expectation maximization algorithm which provides a very good fit. The major drawback is the requirement for data samples. In our system setup the interference depends on the link distance $d$, the road traffic parameters $\lambda,c$, the activity $\xi$ and the channel models. Therefore extensive simulations are needed before regression analysis.

In order to approximate the mean, the variance and the skewness of interference due to transmissions originated behind the transmitter, we may approximate the \ac{PCF} of the hardcore process $\Phi$ by the \ac{PCF} of \ac{PPP} for distance separations larger than $2c$. We have followed the same approach in~\cite{Koufos2018,Koufos2019}, however, without conditioning on the location of a point (the transmitter). The calculation details are available in the supplementary material (optional reading). Over there, we see that the resulting expressions are quite complicated, not insightful about the impact of different parameters on the interference. Because of that, we have also included the expressions after approximating the \ac{PCF} of $\Phi$ by the \ac{PCF} of \ac{PPP} for distance separation larger than $c$. Note that even with this simplification, the correlated locations of vehicles are still retained by the model. Finally, we get
\begin{equation}
\label{eq:Pairs2}
\begin{array}{ccl}
\mathbb{E}^{!o}\!\!\left\{\mathcal{I}\right\} \!\!\!\!\! &\approx& \!\!\!\!\! \displaystyle \frac{\lambda \xi \left(c+d\right)^{1-\eta}}{\eta-1} \\ 
\mathbb{V}^{!o}\!\left\{\mathcal{I}\right\} \!\!\!\!\! &\approx& \!\!\!\!\! \displaystyle \frac{2\lambda\xi\left(c\!+\!d\right)^{1-2\eta}\left(1-\lambda c\xi\right)}{2\eta-1} \\ 
\mathbb{S}^{!o}\!\left\{\mathcal{I}\right\}  \!\!\!\!\! &\approx& \!\!\!\!\! \displaystyle \frac{6\lambda\xi\left(c\!+\!d\right)^{1-3\eta}\left(1\!-\!\lambda c \xi \right)^2}{3\eta-1}\mathbb{V}^{\,!o}\!\left\{\mathcal{I}\right\}^{-3/2}.
\end{array}
\end{equation}

\begin{figure*}[!t]
 \centering 
\subfloat[Mean]{\includegraphics[width=2in]{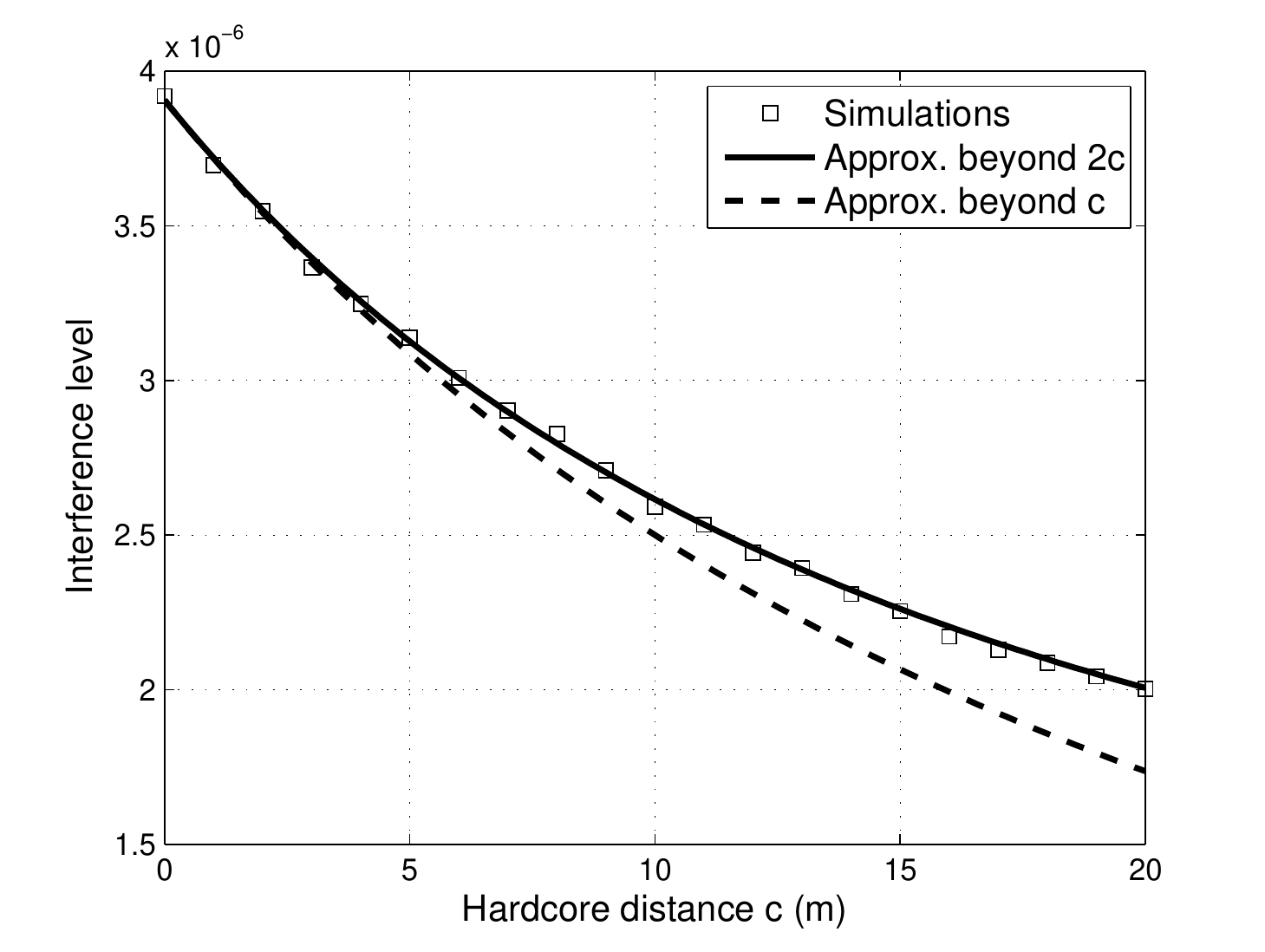}}\hfil 
\subfloat[Standard deviation]{\includegraphics[width=2in]{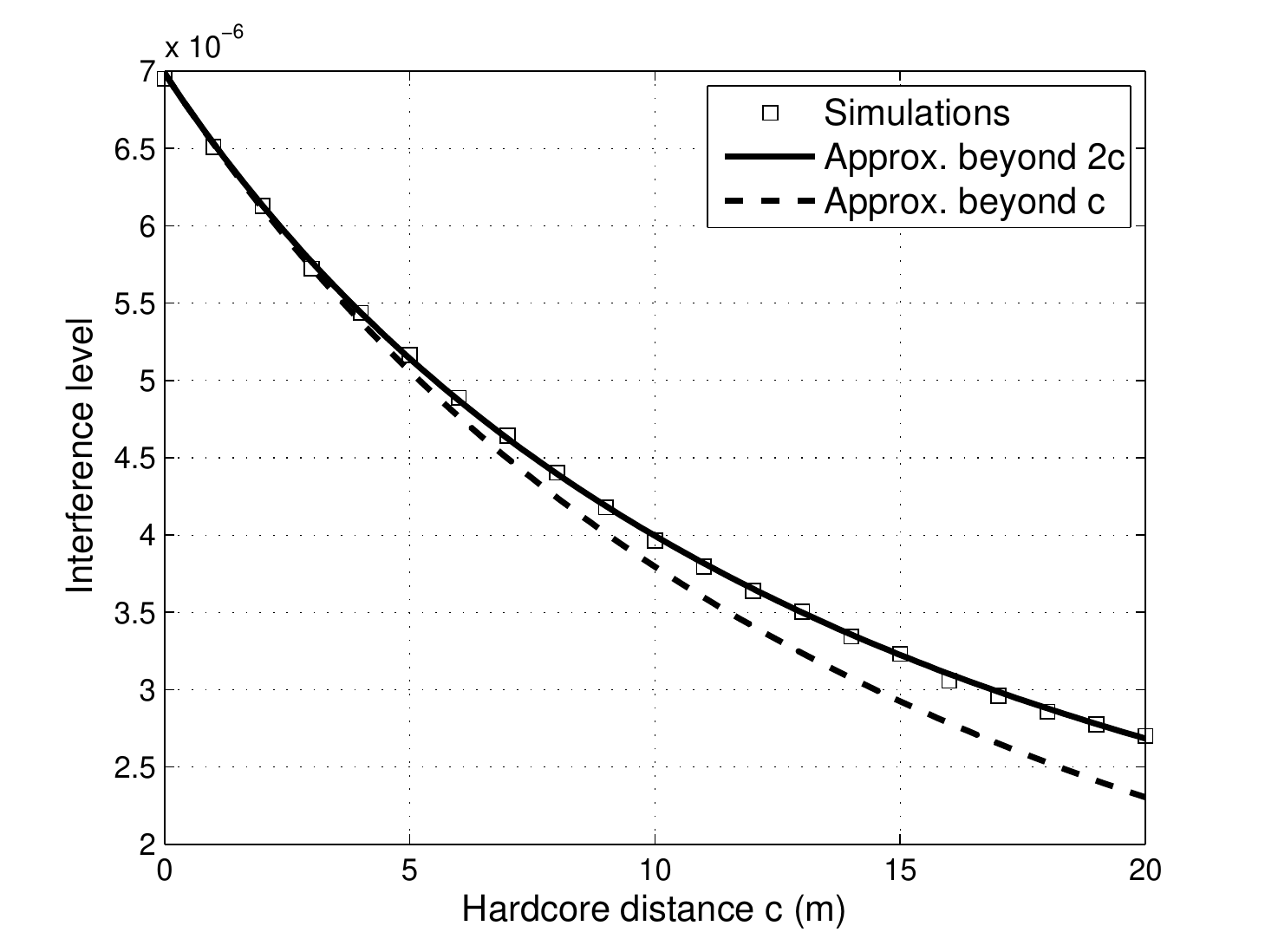}\label{fig:Std}}\hfil \subfloat[Skewness]{\includegraphics[width=2in]{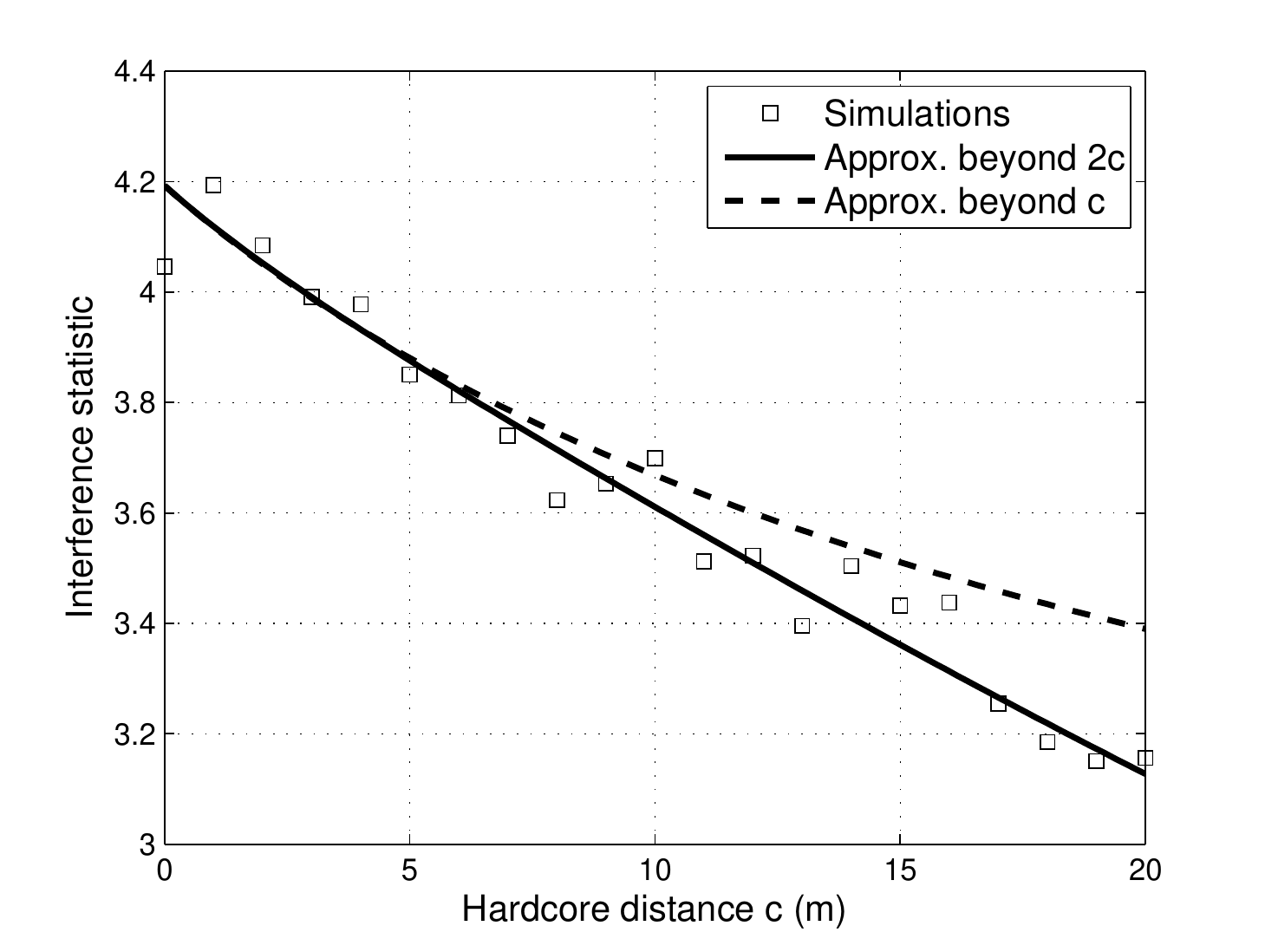}\label{fig:Skewn}}
 \caption{Interference statistics due to transmissions only behind the receiver with respect to the hardcore distance $c$. Intensity of vehicles $\lambda\!=\!0.025 {\text{m}}^{-1}$, pathloss exponent $\eta\!=\!3$, activity probability $\xi\!=\!\frac{1}{2}$, and useful link distance $d\!=\!40 {\text{m}}$. $10^5$ simulations per marker. For the approximations of the statistics using $\rho^{\left(2\right)}\!\left(r\right)\!=\!\lambda^2$ for $r\!>\!c$ see~\eqref{eq:Pairs2}. For the approximations using $\rho^{\left(2\right)}\!\left(r\right)\!=\!\lambda^2$ for $r\!>\!2c$ see the supplementary material.}
 \label{fig:MeanStdSkewn}
\end{figure*}

In Fig.~\ref{fig:MeanStdSkewn} we depict the mean, the standard deviation and the skewness of interference for $\lambda c\!\leq\!\frac{1}{2}$. Approximating the \ac{PCF} for distance separation larger than $2c$ seems very accurate in the estimation of moments. Fortunately, the simpler approximation in~\eqref{eq:Pairs2} captures the general trend in the behavior of interference statistics.

The parameters of the shifted-gamma distribution,  $f_{\mathcal{I}}\!\left(x\right)\!=\! \left(x-\epsilon\right)^{k-1}e^{-\left(x-\epsilon\right)/\beta}/\left(\Gamma\!\left(k\right) \beta^k\right)$, as functions of the link distance $d$ can be estimated, using~\eqref{eq:Pairs2}, as $k\!=\!4/\mathbb{S}^{!o}\!\left\{\mathcal{I}\right\}^2, \beta\!=\!\left(\mathbb{V}^{!o}\!\left\{\mathcal{I}\right\}/k\right)^{1/2}$, and $\epsilon\!=\!\mathbb{E}^{!o}\!\left\{\mathcal{I}\right\}\!-\!k \beta$. The \ac{LT} of interference evaluated at $s\!=\! \theta/P_r\left(d\right)$ is $\mathcal{L}_{\mathcal{I}}\!\left(s\right)\!=\! e^{-s \epsilon}\left(1\!+\!s \beta\right)^{-k}$. 
In order to calculate the moments of interference due to transmissions originated from the vehicles in front of the receiver we follow the same steps as in~\eqref{eq:Pairs2}, remembering to include the attenuation factor $g$ due to antenna backlobes. Finally, we obtain the following approximations:
\begin{equation}
\label{eq:Pairs2b}
\begin{array}{ccl}
\mathbb{E}^{!o}\!\!\left\{\mathcal{I}\right\} \!\!\!\!\! &\approx& \!\!\!\!\! \displaystyle \frac{\lambda \xi g c^{1-\eta}}{\eta-1} \\ 
\mathbb{V}^{!o}\!\left\{\mathcal{I}\right\} \!\!\!\!\! &\approx& \!\!\!\!\! \displaystyle \frac{2\lambda\xi g c^{1-2\eta}\left(1-\lambda c\xi\right)}{2\eta-1} \\ 
\mathbb{S}^{!o}\!\left\{\mathcal{I}\right\}  \!\!\!\!\! &\approx& \!\!\!\!\! \displaystyle \frac{6\lambda\xi g c^{1-3\eta}\left(1\!-\!\lambda c \xi \right)^2}{3\eta-1}\mathbb{V}^{\,!o}\!\left\{\mathcal{I}\right\}^{-3/2}.
\end{array}
\end{equation}

The parameters of the shifted-gamma approximation $\epsilon', \beta', k'$ calculated based on~\eqref{eq:Pairs2b} do not depend on the link distance $d$. Finally, the calculation of the outage probability due to the combined impact of interferers behind and in front of the receiver requires to take the product of the two \acp{LT} and average it over the link distance $d$.
\begin{equation}
\label{eq:Outage}
\begin{array}{ccl}
\mathbb{P}_{\text{out}}\!\left(\theta\right) \!\!\!&=&\!\!\! \displaystyle 1\!-\!\int\nolimits_c^\infty \!\! e^{-\theta\, r^\eta \epsilon\left(r\right)}\! \left(1\!+\!\theta\, r^\eta \beta\!\left(r\right)\right)^{-k\left(r\right)} \times \\  & & \displaystyle e^{-\theta\, r^\eta \epsilon'}\! \left(1\!+\!\theta\, r^\eta \beta'\right)^{-k'} \mu e^{-\mu\left(r-c\right)} {\rm d}r. 
\end{array}
\end{equation}

\begin{figure*}[!t]
 \centering
\subfloat[Own lane only, $L_1$]{\includegraphics[width=2in]{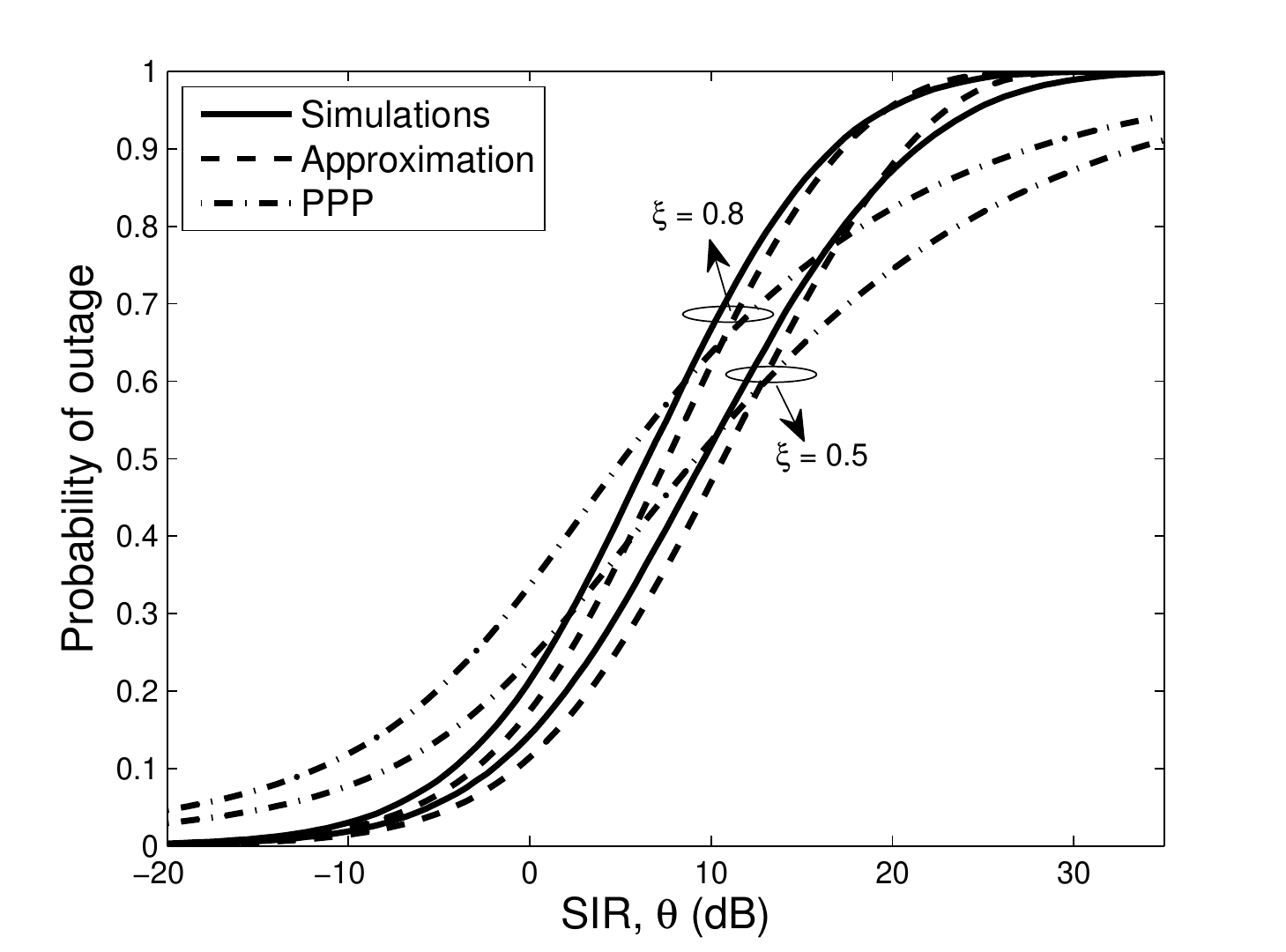}\label{fig:OwnLane}} \hfil
\subfloat[Other lane only, $L_2$]{\includegraphics[width=2in]{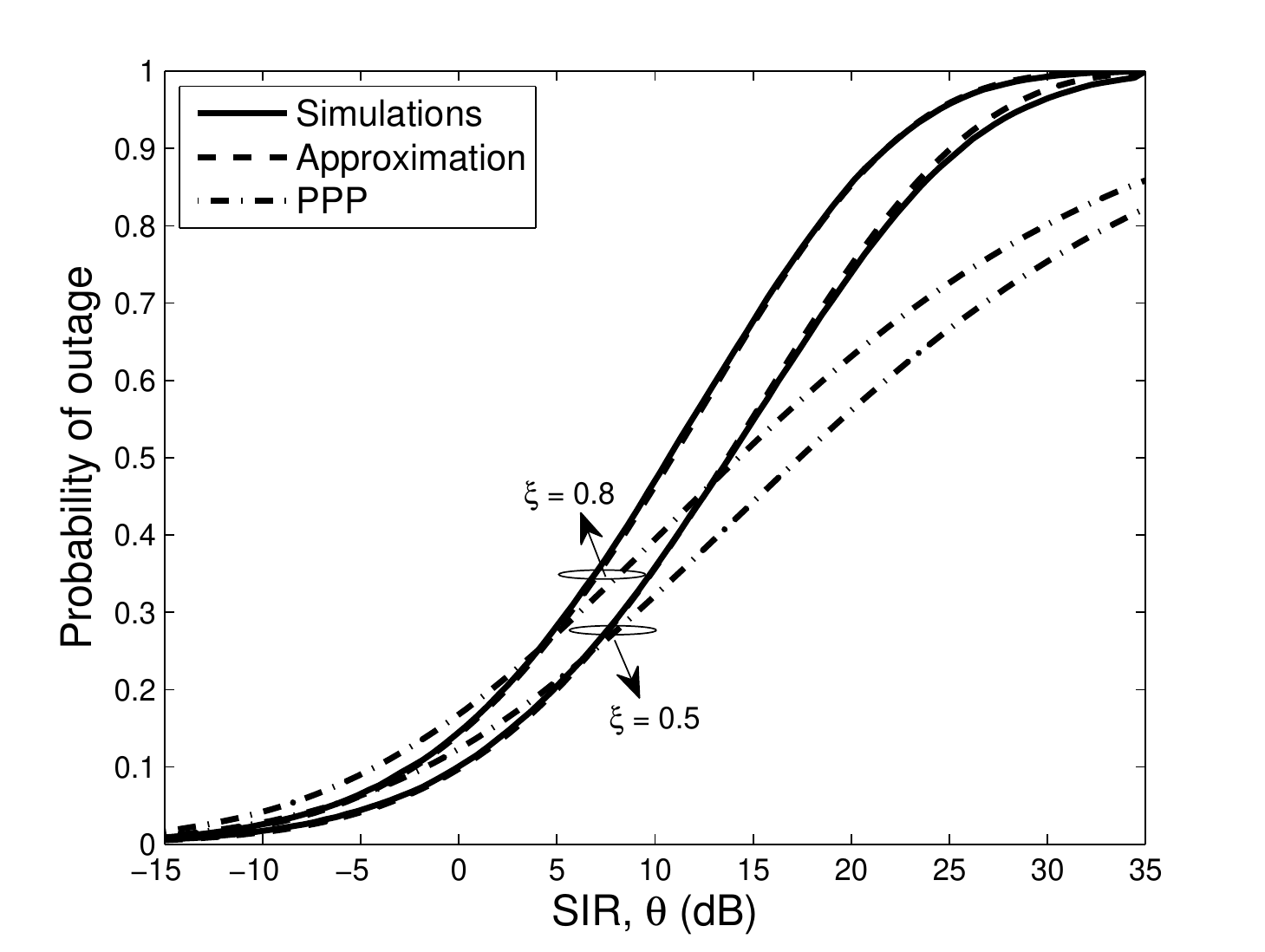}\label{fig:OtherLane}} \hfil 
\subfloat[Both lanes, $L_1 \& L_2$]{\includegraphics[width=2in]{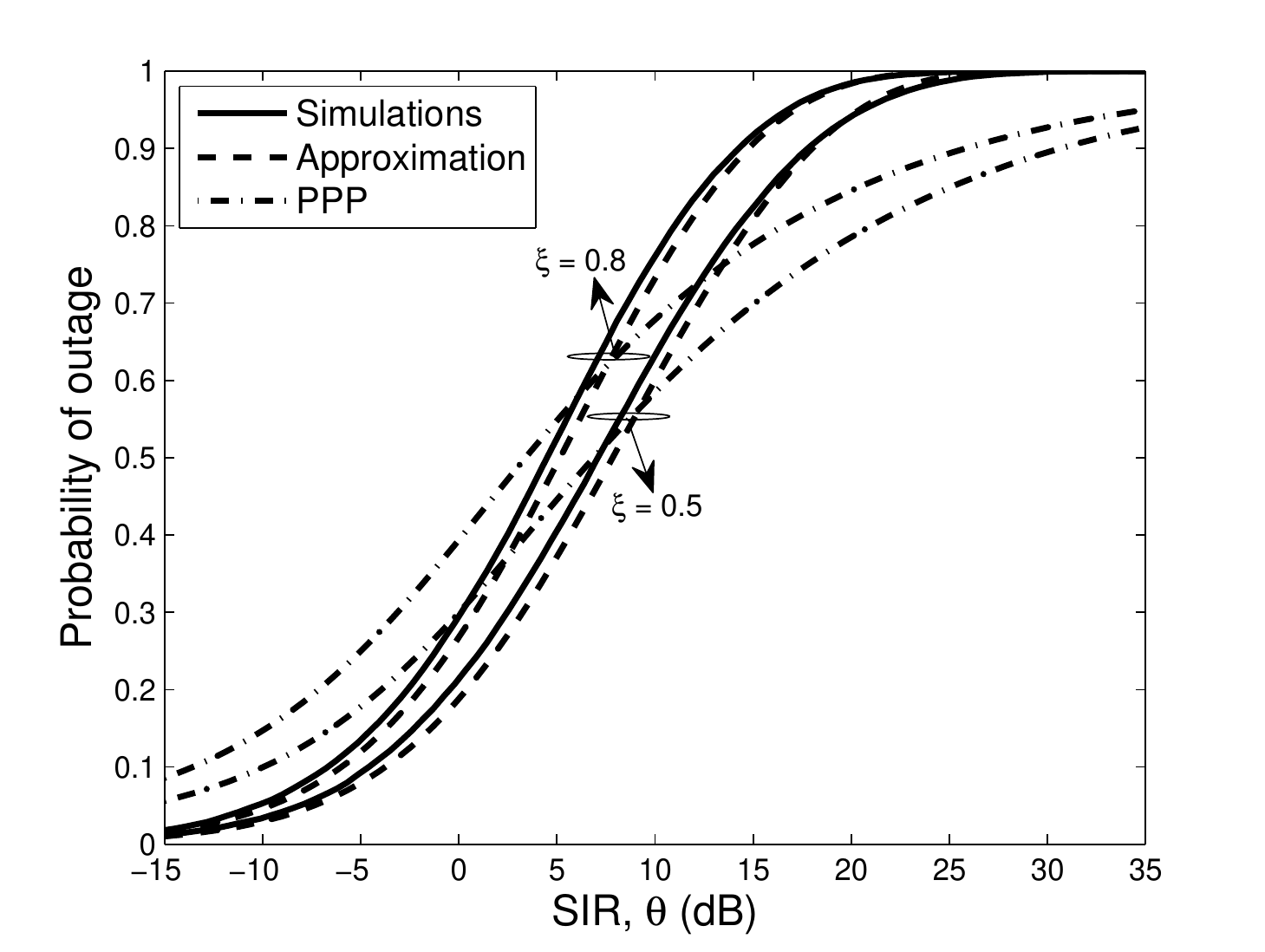}\label{fig:BothLanes}}
 \caption{Probability of outage at the receiver when the interference originates from (a) the lane $L_1$ containing the transmitter-receiver link, (b) another lane $L_2$ parallel to $L_1$, and (c) both lanes. Intensity of vehicles $\lambda\!=\!0.025 {\text{m}}^{-1}$, pathloss exponent $\eta\!=\!3$ and hardcore distance $c\!=\!16 {\text{m}}$. Inter-lane spacing $\ell\!=\!6$ m and antenna beamwidth $\phi\!=\!\frac{\pi}{20}$ give $r_0\!\approx\! 75$ m for $L_2$. $10^5$ simulations. In (a) we have used~\eqref{eq:Outage} to approximate the probability of outage and~\eqref{eq:OutagePPP} to generate the outage probability due to a PPP. In (b) we have used~\eqref{eq:OutageL2} and~\eqref{eq:OutageL2PPP2} respectively. Backlobe antenna attenuation $g\!=\!0.01$.}
\end{figure*}

In Fig.~\ref{fig:OwnLane} we see that the above approximation provides a reasonably good fit to the simulations, while the  \ac{PPP} completely fails. For presentation completeness, the outage probability due to \ac{PPP} has been calculated as 
\begin{eqnarray}
\mathbb{P}_{\text{out}}^{\text{PPP}}\!\left(\theta\right) \!\!\!\!\! &=& \!\!\!\!\! \displaystyle 1\!-\!\!\!\int\nolimits_0^\infty \!\!\!\! e^{\!-\lambda\xi\left( \int\limits_r^\infty \!\! \frac{\theta r^\eta x^{-\eta} {\rm d}x}{1+\theta r^\eta x^{-\eta}} + \int\limits_0^\infty\!\! \frac{g \theta r^\eta x^{-\eta} {\rm d}x}{1+g \theta r^\eta x^{-\eta}} \right)}\lambda e^{-\lambda r} \!{\rm d}r \label{eq:OutagePPPInt} \\ \!\!\!\!\! &=& \!\!\!\!\! \displaystyle 1 \!-\! \frac{\eta\!-\!1}{\left(\eta\!-\!1\right)f\!\left(\theta\right)\!+\!\xi \theta\,  {}_2F_1\!\left(\!1,\!1\!-\!\frac{1}{\eta},\!2\!-\!\frac{1}{\eta},-\theta\!\right)},
\label{eq:OutagePPP}
\end{eqnarray}
where $f\!\left(\theta\right)\!=\!1\!+\!\frac{\pi}{\eta}\csc\left(\frac{\pi}{\eta}\right)\xi\left(g \theta\right)^{\frac{1}{\eta}}$.
 
It is worth to mention that the outage probability in~\eqref{eq:OutagePPP} does not depend on the intensity $\lambda$. This resembles the calculation of downlink coverage probability in \ac{PPP} cellular networks, which is also independent of the intensity of base stations in the interference-limited regime with nearest base station association~\cite[equation (14)]{Andrews2011}. For the hardcore process, the outage probability in~\eqref{eq:Outage} depends on the intensity $\lambda$ through the parameter $\mu$ and also through the parameters of the shifted-gamma distribution, see~\eqref{eq:Pairs2} and~\eqref{eq:Pairs2b}. 

\section{Multi-lane motorway \acp{VANET}}
\label{sec:MultiLane}
\begin{figure}[!t]
 \centering
  \includegraphics[width=\columnwidth]{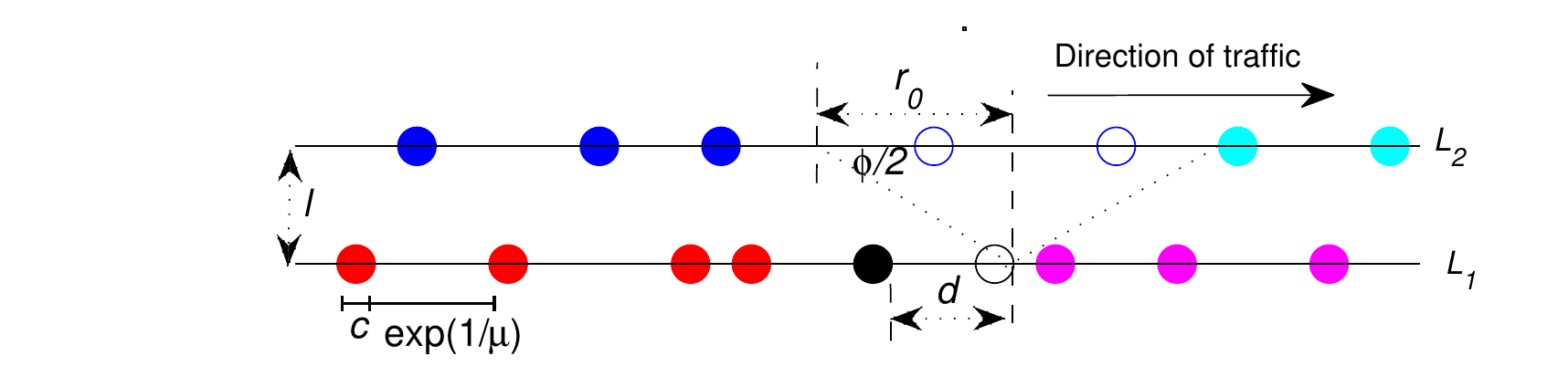}
 \caption{The vehicles of lane $L_2$ at distances larger than $r_0$ behind (blue disks) and in front of (cyan disks) the receiver are potential sources of interference. We ignore the interference from vehicles due to antenna sidelobe radiation (hollow blue disks). The receiver (hollow black disks) at lane $L_1$ is placed at the origin to facilitate the interference analysis due to vehicles at lane $L_2$. See also the caption of Fig.~\ref{fig:SystModelBeam}.}
 \label{fig:SystModelBeam2}
\end{figure}

Let us consider another lane, $L_2$, with same road traffic parameters $\lambda,c$ and same activity probability $\xi$ for the vehicles. Extension to more lanes is straightforward, and numerical examples will be given in the next section. The link under consideration is still located at $L_1$; $L_2$ is just an extra source of interference. Let us denote by $\ell$ the inter-lane separation and by $\phi$ the antenna beamwidth. Then, the interfering vehicles from $L_2$ are located at distances larger than $r_0\!=\!\ell/\tan\frac{\phi}{2}$ from the receiver, see Fig.~\ref{fig:SystModelBeam2} for an illustration. Note that the guard zone $r_0$ is expected to be much larger than the lane separation $\ell$ for practical values of $\phi$ and $\ell$. Because of that, we may neglect the impact of $\ell$ in the distance-based pathloss of other-lane interference without introducing much error as compared to the simulations. 

Unlike $L_1$, the interference analysis for $L_2$ does not require conditioning. Keeping in mind that the point process is stationary, the mean interference due to $L_2$ does not depend on the correlation properties but only on the intensity $\lambda$. The variance of interference due to $L_2$ has been calculated in~\cite[Equation (14)]{Koufos2018} and the skewness in~\cite[Lemma 1]{Koufos2019}. After scaling the moments to consider the attenuation $g$ for the transmissions of the vehicles ahead of the receiver, the approximations become 
\begin{equation}
\label{eq:MomOtherLane}
\begin{array}{ccl}
\mathbb{E}\!\left\{\mathcal{I}\right\} \!\!\!\!\! &=& \!\!\!\!\! \displaystyle \frac{\lambda\xi\left(1+g\right) r_0^{1-\eta}}{\eta-1} \\ 
\mathbb{V}\!\left\{\mathcal{I}\right\} \!\!\!\!\! &\approx& \!\!\!\!\! \displaystyle \frac{2\lambda\xi \left(1+g\right) r_0^{1-2\eta}}{2\eta-1}\left(1\!-\!\lambda c \xi \!+\! \frac{1}{2}\lambda^2 c^2 \xi^2 \right) \\
\mathbb{S}\!\left\{\mathcal{I}\right\} \!\!\!\!\! &\approx& \!\!\!\!\! \displaystyle \frac{6 \lambda\xi  r_0^{1-3\eta}}{\left(3\eta\!-\!1\right)\sqrt{1\!+\!g}} \! \left( \frac{2\lambda\xi r_0^{1-2\eta}}{2\eta-1}\right)^{\!\!-\frac{3}{2}}\!\!\!\left(1\!-\!\frac{\lambda c \xi}{2}\right)\!.
\end{array}
\end{equation}

We will use a shifted-gamma approximation for the distribution of interference from $L_2$. The moments of interference and subsequently the gamma parameters for $L_2$, $k,\beta,\epsilon$, do not depend on the link distance $d$ as in~\eqref{eq:Pairs2} and~\eqref{eq:Outage}. The outage probability due to $L_2$ (only) can be calculated by integrating the \ac{LT} of interference over the link distance. 
\begin{eqnarray}
\mathbb{P}_{\text{out}}\!\left(\theta\right) \!\!\!\!\! &=& \!\!\!\!\! \displaystyle 1\!-\!\int\nolimits_c^\infty \!\!\!\! e^{-\theta\, r^\eta \epsilon}\left(1\!+\!\theta\, r^\eta \beta\right)^{-k}\mu e^{-\mu\left(r-c\right)} {\rm d}r \label{eq:OutageL2} \\ \mathbb{P}_{\text{out}}^{\text{PPP}}\!\left(\theta\right) \!\!\!\!\! &=& \!\!\!\!\! \displaystyle 1 \!\!-\!\! \int_0^\infty \!\!\!\! e^{- \lambda \xi \left(1+g\right) \int_{r_0}^\infty \frac{\theta r^\eta x^{-\eta}}{1+\theta r^\eta x^{-\eta}}{\rm d}x }\! \lambda e^{-\lambda r} \! {\rm d}r  \label{eq:OutageL2PPP} \\ \!\!\!\!\! &=& \!\!\!\!\! \displaystyle 1 \!\!-\!\! \int_0^\infty \!\!\!\! e^{- \lambda \xi \left(1+g\right) t\left(r,r_0,\theta\right)}\! \lambda e^{-\lambda r} \! {\rm d}r, \label{eq:OutageL2PPP2}
\end{eqnarray}
where $t\!\left(r,r_0,\theta\right)\!=\!\frac{\pi}{\eta}\csc\left(\frac{\pi}{\eta}\right)\theta^{\frac{1}{\eta}}r-r_0\, {}_2F_1\left(1,\frac{1}{\eta}\,1\!+\!\frac{1}{\eta},\!-\frac{r_0^\eta}{\theta r^n}\right)$.

Note that the function $t\!\left(\cdot\right)$ does not depend linearly on $r$. This is because the lower integration limit with respect to $x$ in~\eqref{eq:OutageL2PPP} is the constant $r_0$ and not $r$ as in~\eqref{eq:OutagePPPInt}. Because of that, a closed-form calculation of~\eqref{eq:OutageL2PPP2} is not possible. In Fig.~\ref{fig:OtherLane} we have validated~\eqref{eq:OutageL2} and~\eqref{eq:OutageL2PPP2} against simulations. In order to calculate the outage probability under aggregate interference from $L_1$ and $L_2$, we need to multiply the \acp{LT} of interference from the two lanes before integrating over the link distance. The results are shown in Fig.~\ref{fig:BothLanes}. 

\section{Probability of outage $-$ synthetic traces}
\label{sec:MultiLaneSynthetic}
For the synthetic traces, we denote by $L_3$ the left lane characterized by the highest average speed, by $L_2$ the middle lane and by $L_1$ the right lane. We will carry out $10^5$ simulation runs for the outage probability per snapshot. For each of them, we construct first the empirical \ac{CDF} of inter-vehicle distances. In each simulation run we sample it (using linear interpolation), generating enough samples to cover a roadway of $10$ km. The link is located in the middle lane, and the link distance $d$ is generated by sampling the empirical \ac{CDF} of $L_2$. Independent samples for the fading and the activity for each vehicle are also generated. Finally, the interference levels from the three lanes are aggregated at the receiver.

For the outage probability predictions using the hardcore point process, we use the least-square estimates $\hat{\lambda}_i, \hat{c}_i$, see Section~\ref{sec:Validation}, and generate the \ac{LT} of interference for each lane, $L_{\mathcal{I}_i}\!\left(\theta,r\right)$. Then, we integrate the product of \acp{LT} over a shifted-exponential distribution with parameters $\hat{\mu}_2,\hat{c}_2$.
\begin{equation}
\label{eq:PoutModelFinal}
\mathbb{P}_{\text{out}}\!\left(\theta\right) \!=\! 1\!\!-\!\!\int\nolimits_{\hat{c}_2}^\infty \prod\nolimits_{j=1}^3 L_{\mathcal{I}_j}\!\!\left(\theta,r\right) \hat{\mu}_2 e^{-\hat{\mu}_2\left(r-\hat{c}_2\right)} {\rm d}r, 
\end{equation}
where $L_{\mathcal{I}_j}\!\left(\theta,r\right)\!=\!e^{-\theta\, r^\eta \epsilon_j}\!\left(1\!+\!\theta r^\eta \beta_j\right)^{-k_j}, j\!\in\!\left\{1,3\right\}$ and $L_{\mathcal{I}_2}\!\left(\theta,r\right)\!=\!e^{\!-\theta\, r^\eta \left( \epsilon_2\left(r\right) + \epsilon_2'\right)}\!\left(1\!+\!\theta\, r^\eta \beta_2\!\left(r\right)\right)^{\!-k_2\left(r\right)}\!\! \left(1\!+\!\theta\, r^\eta \beta_2'\right)^{\!-k_2'}$. The parameters $k_j,\beta_j,\epsilon_j$ do not depend on $r$, and they are calculated via moment matching using the estimates $\hat{\lambda}_j,\hat{c}_j$ in~\eqref{eq:MomOtherLane}. The parameters $k_2\!\left(r\right),\beta_2\!\left(r\right),\epsilon_2\!\left(r\right)$ are calculated by the \ac{MoM} using the estimates $\hat{\lambda}_2,\hat{c}_2$ in~\eqref{eq:Pairs2}, and the parameters $k_2',\beta_2',\epsilon_2'$ are calculated using $\hat{\lambda}_2,\hat{c}_2$ in~\eqref{eq:Pairs2b}. 

For the outage probability predictions using \ac{PPP}, we use the \ac{MLE} $\hat{\lambda}_i$ from the data sample, and we integrate the product of \acp{LT} over an exponential distribution with parameter $\hat{\lambda}_2$. 
\begin{equation}
\label{eq:PoutPPPFinal}
\mathbb{P}_{\text{out}}^{\text{PPP}}\!\left(\theta\right) \!=\! 1\!\!-\!\!\int\nolimits_0^\infty \prod\nolimits_{j=1}^3 L_{\mathcal{I}_j}^{\text{PPP}}\!\!\left(\theta,r\right) \hat{\lambda}_2 e^{-\hat{\lambda}_2 r} {\rm d}r, 
\end{equation}
where $L_{\mathcal{I}_2}^{\text{PPP}}\!\left(\theta,r\right)\!\!=\!\! e^{-\hat{\lambda}_2\xi r \left( \frac{\theta}{\eta-1}{}_2F_1\!\left(1,1-\frac{1}{\eta},2-\frac{1}{\eta},-\theta\right) + f\left(\theta\right)-1\right)}$ and $L_{\mathcal{I}_j}^{\text{PPP}}\!\left(\theta,r\right)\!=\!e^{-\hat{\lambda}_j \xi \left(1+g\right) t\left(r,r_0,\theta\right)},\, j\!\in\!\left\{1,3\right\}$. 

\begin{figure}[!t]
 \centering
{\includegraphics[width=3in]{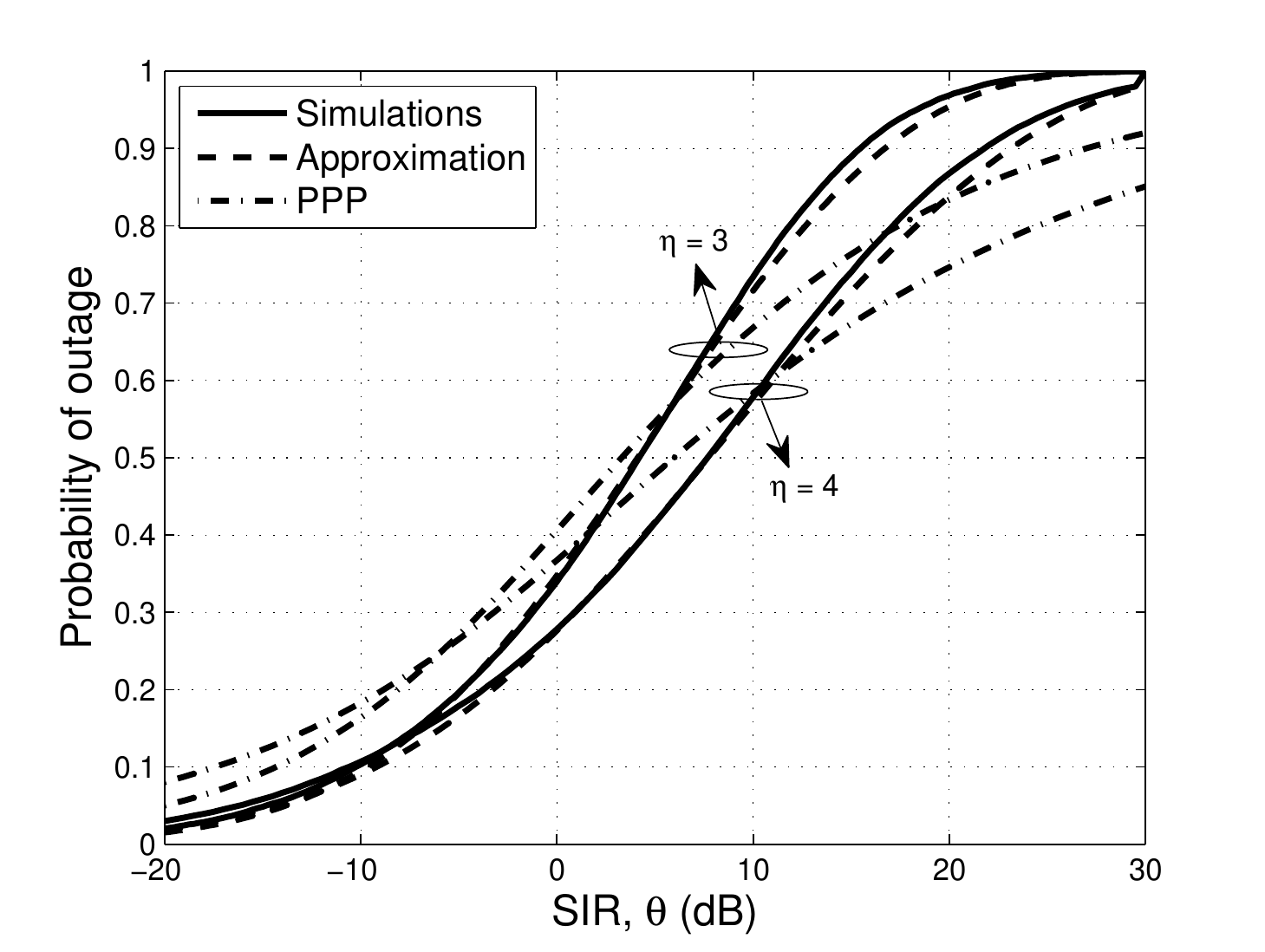}}
 \caption{Probability of outage due to interference originated from three lanes with synthetic traces. $10^5$ simulation runs. Activity $\xi\!=\!\frac{1}{2}$, inter-lane spacing $\ell\!=\!4$ m and antenna beamwidth $\phi\!=\!\frac{\pi}{20}$ yield $r_0\!\approx\! 50$ m for $L_1$ and $L_3$, $g\!=\!0.01$.}
 \label{fig:ThreeLanes}
\end{figure}

For illustration purposes, we select the 1000-th snapshot. In the busy hour, we estimate $\hat{\lambda}_1\!=\!0.0248, \hat{\lambda}_2\!=\!0.0218$ and $\hat{\lambda}_3\!=\!0.0205$, and $\hat{c}_1\!=\!7.10, \hat{c}_2\!=\!11.05$ and $\hat{c}_3\!=\!14.82$ for the hardcore processes. In Fig.~\ref{fig:ThreeLanes}, we see that equation~\eqref{eq:PoutModelFinal} predicts very well the simulated outage probability using the sampled point set. On the other hand, the \ac{PPP} prediction using~\eqref{eq:PoutPPPFinal} with estimates $\hat{\lambda}_1\!=\!0.0215, \hat{\lambda}_2\!=\!0.0196$ and $\hat{\lambda}_3\!=\!0.0195$ is poor. In addition, we illustrate that with higher pathloss exponent, the resulting lower interference level dominates over the lower received signal level, and the outage probability decreases.  

\begin{figure*}[!t]
 \centering
\subfloat[motorway M40]{\includegraphics[width=3in]{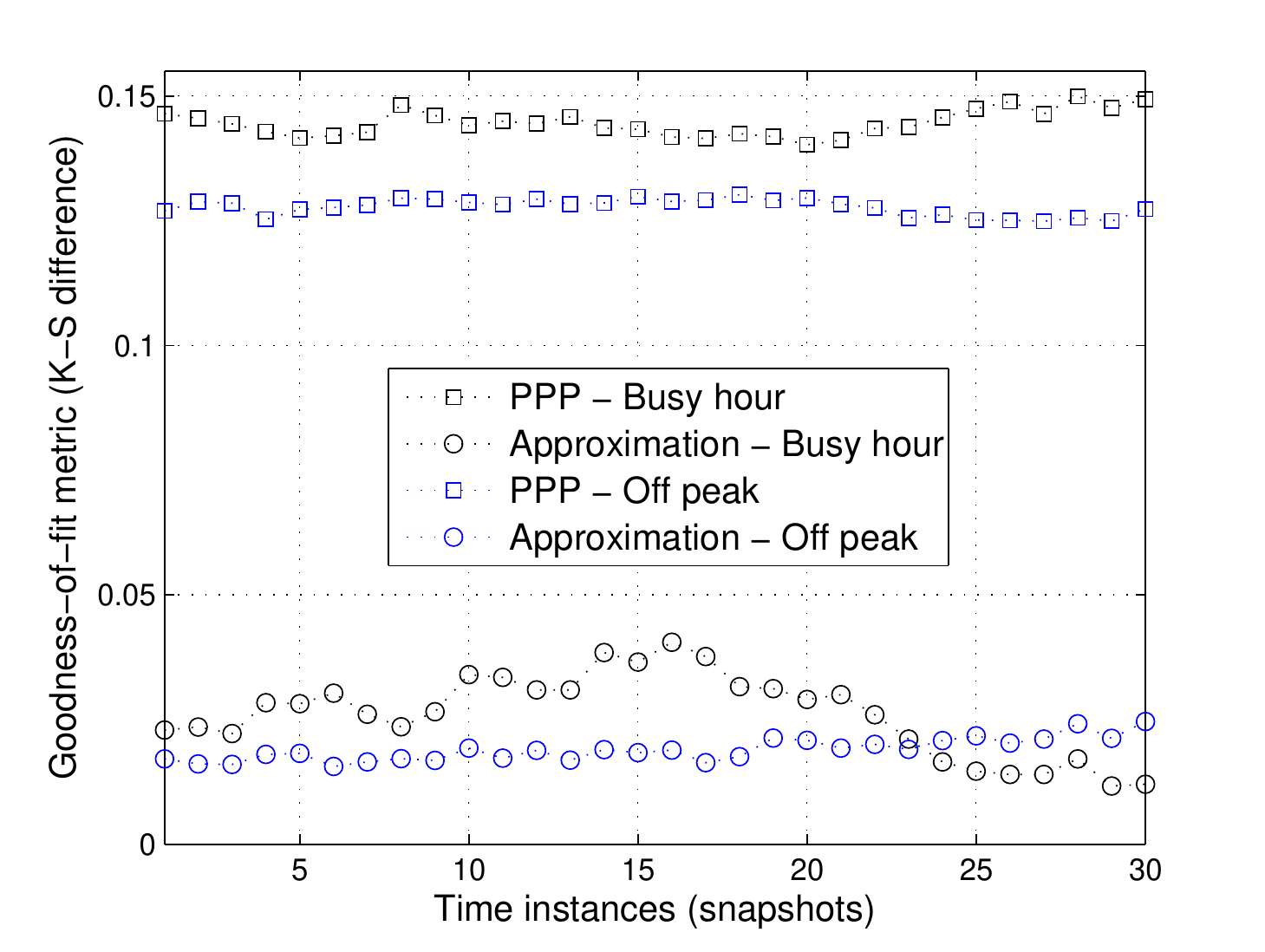}\label{fig:MetricsBusy}} 
\subfloat[motorway A6]{\includegraphics[width=3in]{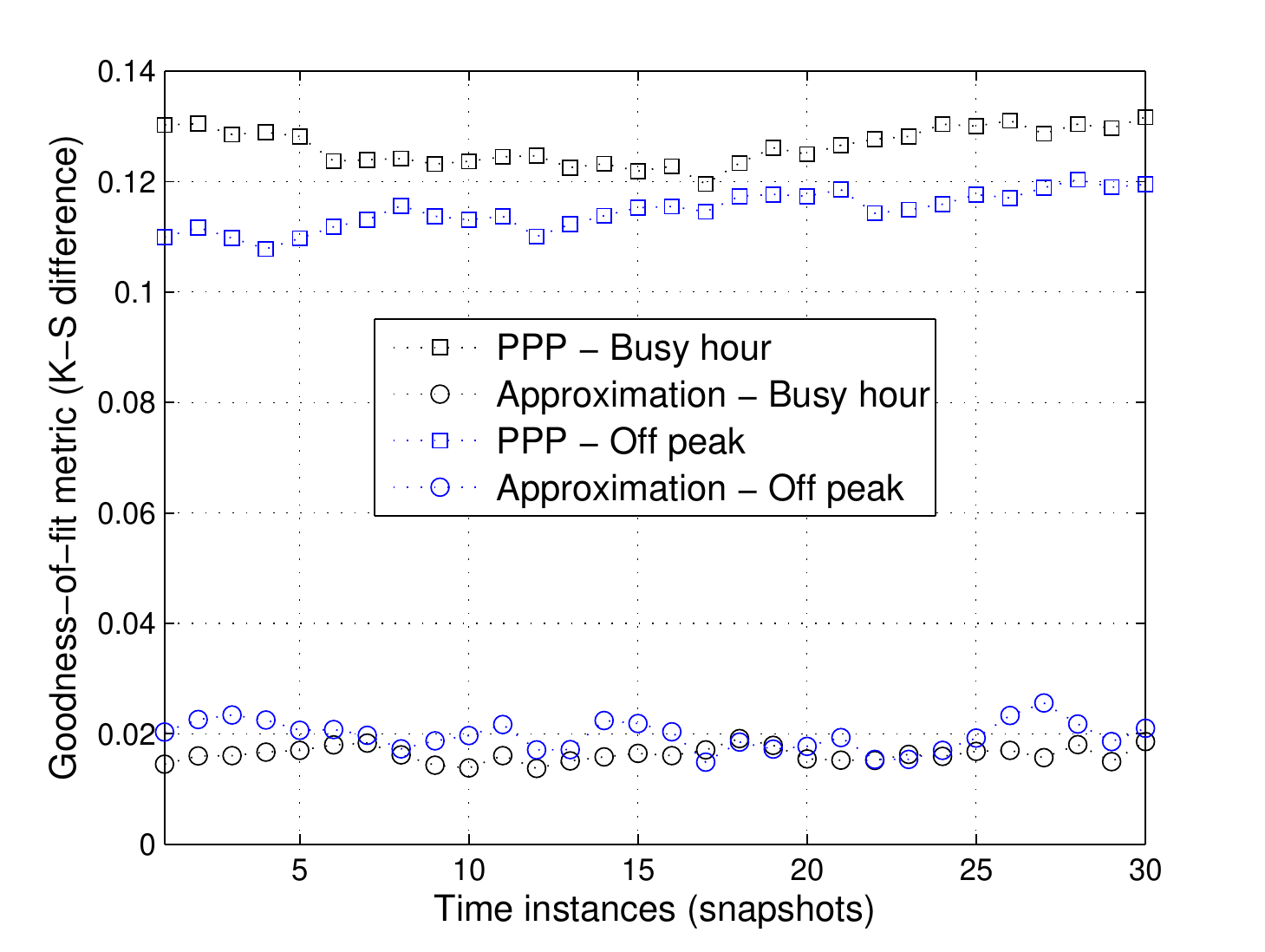}\label{fig:MetricsOffA6}}
 \caption{Goodness-of-fit for the approximations of the empirical outage probability using~\eqref{eq:PoutModelFinal} and~\eqref{eq:PoutPPPFinal} for $60$ snapshots ($30$ during off-peak and $30$ during the busy hour) and two motorways. Activity $\xi\!=\!\frac{1}{2}$, pathloss exponent $\eta\!=\!3$, guard zone $r_0\!=\!75$ m, backobe antenna attenuation $g\!=\!0.01$, and $10^5$ simulation runs per snapshot.}
 \label{fig:Metrics}
\end{figure*}

In order to assess the quality of the approximations, we also calculate the maximum vertical difference between the simulated \ac{CDF} and the \acp{CDF} obtained from the two models, which is the metric used in Kolmogorov-Smirnov test. In Fig.~\ref{fig:Metrics}, we depict the goodness-of-fit for two motorways M40 and A6 outside Madrid collected at the same day and time. The motorway A6 exhibits higher (lower) traffic intensity than M40 during the busy hour (off-peak), see~\cite[Table II]{Fiore2014} for a comparison of the average road traffic characteristics between M40 and A6. We see that our model consistently gives much closer predictions to the empirical \ac{CDF} than the \ac{PPP}. 

The estimates $\hat{\lambda}_i,\hat{c}_i$ for the hardcore point process could have also been obtained by directly fitting them to minimize the square difference between the estimated and empirical outage probabilities. This is known as the method of minimum contrast, see~\cite[Section IV]{Guo2013} for fitting the Strauss and Poisson hardcore processes to snapshots of macro base stations. Note that this method would require extensive numerical search in our case, because there are six parameters $\lambda_i,c_i, i\!\in\!\left\{1,2,3\right\}$ to fit, and the outage probability is a complicated function of them. On the other hand, estimating $\lambda_i,c_i$ from the trace, and plugging them into the interference model has much lower complexity, and it still provides a very good estimate for the outage probability.  

\section{Conclusions}
\label{sec:Conclusions}
In this paper we developed a low-complexity model for the probability of outage in multi-lane \acp{VANET} in realistic enironments~\cite{Fiore2014,Fiore2015}. The model consists of two parts: Firstly, it borrows from transportation (and statistical mechanics) literature a simple extension (with a hardcore distance) to the \ac{PPP} for the deployment of vehicles along a lane. This does not come without cost because introducing hardcore distance makes the locations of vehicles correlated. Secondly, it applies the method of moments using a shifted-gamma distribution to approximate the Laplace transform of generated interference per lane. We constructed simple but accurate approximations for the first three moments of interference under Rayleigh fading with and without conditioning. We have seen very good prediction of the outage probability in realistic motorway setups, while the \ac{PPP} fails. Instead of running time-consuming simulations, the system designer may estimate the intensity of vehicles and the hardcore distance from the available trace, and use a single numerical integration, see~\eqref{eq:PoutModelFinal}, to assess the probability of outage. The model should be particularly useful in cases with high transmission probability because, with strong thinning, the hardcore process converges to \ac{PPP}~\cite{Hourani2018}. Potential direction for future work is the application of more realistic propagation functions and fading channels for \ac{V2V} communication. In addition, while a hardcore process has fitted well the available motorway traces, the development of point processes tailored to strong clustering of vehicles might be needed to model urban traffic conditions. Network-wide performance evaluation along the motorway using meta distributions is also an important direction for future work~\cite{Koufos2019b}. 

\section*{Acknowledgment}
This work was supported by the EPSRC grant number EP/N002458/1 for the project Spatially Embedded Networks. All underlying data are provided in full within this paper.

\end{document}